\documentclass[11pt]{article}

\usepackage[a4paper,margin=2.2cm]{geometry}
\usepackage[T1]{fontenc}
\usepackage[utf8]{inputenc}
\usepackage{lmodern}
\usepackage{microtype}
\usepackage{amsmath,amssymb,mathtools,bm}
\usepackage{graphicx}
\usepackage{booktabs}
\usepackage{algorithm}
\usepackage{algpseudocode}
\usepackage{longtable,tabularx}
\usepackage{array}
\usepackage{xcolor}
\usepackage{caption}
\usepackage{lineno}
\usepackage{setspace}
\usepackage{cite}
\usepackage{hyperref}
\usepackage{cleveref}
\usepackage{placeins}
\usepackage{titlesec}

\graphicspath{{figures/}{./}}
\hypersetup{
    colorlinks=true,
    linkcolor=black,
    citecolor=black,
    urlcolor=blue,
    pdfauthor={Guillermo Herrera Sanchez, Daniel Gradeci, Daniele Ramsay, Safe Khan},
    pdftitle={The life and death of football team runs}
}

\titleformat{name=\section,numberless}
  {\normalfont\large\bfseries}{}{0pt}{}
\titleformat{name=\subsection,numberless}
  {\normalfont\normalsize\bfseries}{}{0pt}{}
\titlespacing*{\section}{0pt}{2.0ex plus 0.5ex minus 0.2ex}{0.8ex}
\titlespacing*{\subsection}{0pt}{1.5ex plus 0.4ex minus 0.2ex}{0.5ex}

\newcommand{\dd}{\mathrm{d}}
\newcommand{\Prob}{\mathbb{P}}
\newcommand{\Expect}{\mathbb{E}}
\newcommand{\FigRef}[1]{Fig.~\ref{#1}}
\newcommand{\EqRef}[1]{Eq.~\eqref{#1}}
\newcommand{\FullWidthFigure}[1]{%
    \noindent\makebox[\textwidth][c]{%
        \includegraphics[width=\textwidth,keepaspectratio]{#1}%
    }%
}
\newcommand{\positivepart}[1]{\left(#1\right)_{+}}

\title{
\textbf{
The life and death of football team runs:
survival of collective modes shapes L\'evy-like transport
}
}

\author{
Guillermo Herrera Sánchez,
Daniel Gradeci,
Daniele Ramsay,
Safe Khan$^{*}$\\[0.5em]
\small Framewave Lab, London, United Kingdom\\
\small $^{*}$contact@framewave-lab.com
}

\date{}

\begin{document}
\maketitle

\begin{abstract}
Broad run-length distributions and short-lag superdiffusion have recently
been reported in football players and team centroids, prompting a
collective-foraging interpretation. The mechanism linking these player- and
team-level signatures remains unclear. Using SoccerMon GPS data from 66
tracked team-match records across 62 fixtures in the Norwegian women's
top-flight Toppserien, we asked whether player transport is inherited from a
translating team mode and what controls the lifetime of that mode. We
decomposed player displacement into centroid translation and motion within
the formation, then modelled the switching and termination of directional
centroid runs. At longer lags, centroid translation carried an increasing
share of player displacement. Run lifetimes were broad but finite:
termination was highest near onset, declined with age and rose modestly
later. After accounting for speed and directional persistence, collective
order added only limited predictive information. At the population level, an age-dependent switching-and-termination model
predicted, on held-out match dates, the early enrichment and later depletion
of high-order states among surviving runs. These results link player- and team-level L\'evy-like movement
through a finite-lived collective mode whose age and internal state shape its
survival. Long displacements need not be selected in advance; they can emerge
when transient collective modes persist, reorganise and selectively survive.
More broadly, trajectory statistics may record which dynamical histories
survive, not only the rules by which agents move.
\end{abstract}


\section*{Introduction}

L\'evy walks are finite-speed trajectories composed of persistent runs whose
durations or lengths span a broad range of scales. When run times are
sufficiently broad, such motion can generate superdiffusion over intermediate
times \cite{Viswanathan1999,Zaburdaev2015}. L\'evy-like movement has been
reported in microorganisms, insects, marine predators, birds and human
mobility, and is often interpreted through theories of efficient search or
foraging \cite{Viswanathan1999,Edwards2007,Zaburdaev2015}.

A L\'evy-like trajectory does not identify a unique statistical family. In
finite samples, truncated power laws, log-normal distributions and other broad
forms can be difficult to distinguish \cite{Edwards2007}. Nor does the
trajectory identify a unique mechanism: similar statistics can arise from
environmental heterogeneity, behavioural switching, interactions,
multiplicative fluctuations or population mixtures
\cite{ReynoldsOuellette2016,FedotovKorabel2017}. Residence-time-dependent
turning can broaden run lifetimes, whereas random death or other termination
processes can temper them \cite{FedotovTanZubarev2015,Stage2017}. The
mechanistic question is therefore not only which distribution describes
completed movements, but why some movements survive far longer than others.

Football provides repeated, densely tracked realisations of a collective
moving under shared tactical and spatial constraints. Players coordinate with
teammates, respond to opponents and continually reorganise within a finite
pitch, so each match contains many examples of collective movements forming,
changing and ending. Previous studies have quantified team centroids, occupied
areas, formations, pairwise coordination and transitions among collective
states \cite{Frencken2011,Bartlett2012,Moura2013,Welch2021,Marcelino2020},
while stochastic models have explored how player interactions generate
team-level dynamics \cite{Chacoma2021}. More recently, broad run lengths and
short-lag superdiffusion were reported in both individual-player and
team-centroid trajectories, motivating a collective-foraging interpretation
\cite{Shpurov2024}.

These findings raise two questions. First, are the player- and team-level
transport signatures independent, or is longer-timescale player motion carried
by a translating collective mode? Second, what determines the lifetime of
that mode? Here, a translating collective mode is a directional run of the
team centroid: a period during which the centroid continues along a similar
heading. A distribution of completed steps cannot answer either question
because it neither separates translation of the team from rearrangement within
the formation nor reveals how termination changes as a run ages.

We analysed SoccerMon GPS tracking from Rosenborg and V\aa lerenga across 62
competitive fixtures in the 2020 and 2021 Norwegian women's top-flight
Toppserien seasons, comprising 66 tracked team-match records. We first
decomposed player displacement into translation of the team centroid and
motion relative to the formation. We then segmented the centroid trajectory
into directional runs and modelled each run as a finite-lived collective
transport state. Dividing each run into one-second intervals allowed us to
relate termination to run age and preceding order. Grouped
leave-one-match-date-out cross-fitting withheld all tracked records played on
the same date. Finally, age-dependent transitions among low-, mid- and
high-order states were used to predict the composition of the surviving run
population.

These analyses form a single chain: the displacement decomposition establishes
centroid runs as the relevant collective transport object; the termination
model identifies how their lifetimes depend on age and order; and the state
model explains how switching and differential termination reshape the
population that survives. Empirically, centroid translation accounted for an
increasing share of player displacement at longer lags, termination depended
strongly on run age and more weakly on preceding order, and the state-structured
model reproduced the changing order composition of surviving runs. Together,
these results support a survival-and-reorganisation mechanism for broad
collective transport.

\begin{figure}[!htbp]
    \centering
    \FullWidthFigure{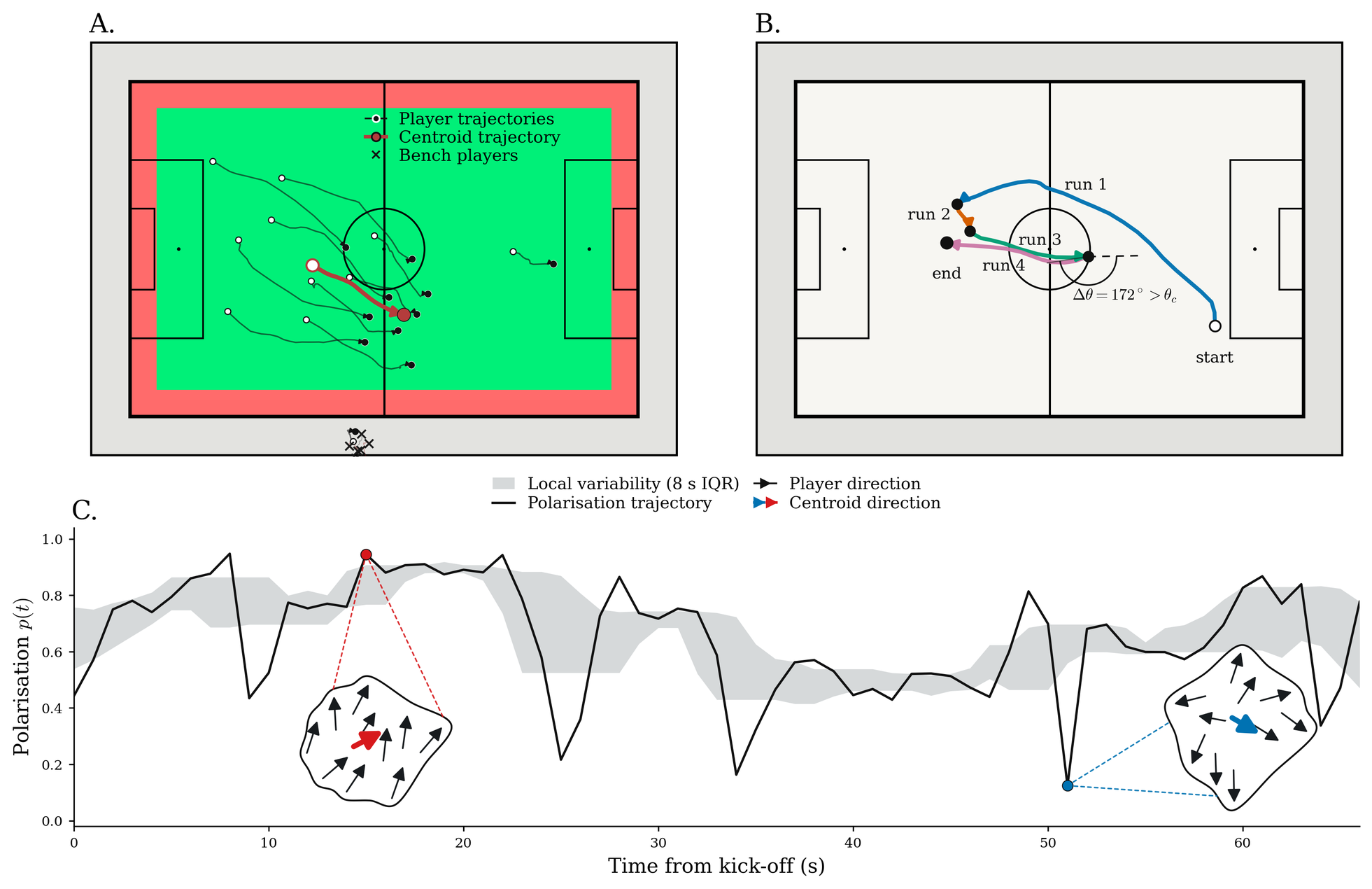}
    \caption{
    \textbf{From player tracking to directional collective runs.}
    \textbf{A}, Illustrative match interval showing active-player trajectories
    and the corresponding team-centroid trajectory. Excluded bench and off-pitch tracks are shown separately.
    \textbf{B}, Segmentation of the centroid path into directional runs. A new
    run begins when the change between consecutive centroid movement directions
    exceeds $\theta_c=30^{\circ}$.
    \textbf{C}, Time-resolved collective polarisation $p(t)$ during an
    illustrative match interval. The grey band shows the rolling 8-s
    interquartile range. Insets contrast a highly polarised translating
    configuration with a weakly polarised configuration.
    }
    \label{fig:framework}
\end{figure}

\FloatBarrier

\section*{Results}

\subsection*{Player transport is increasingly carried by collective translation}

We first asked whether longer-timescale player displacement reflects
translation of the team or rearrangement within the formation. To separate
these contributions, the pitch-frame position of active player $i$ was
decomposed exactly as
\begin{equation}
    \bm x_i(t)
    =
    \bm X_{\mathrm c}(t)
    +
    \bm r_i(t),
    \label{eq:position_decomposition}
\end{equation}
where $\bm X_{\mathrm c}(t)$ is the team centroid and $\bm r_i(t)$ is the
player's position relative to that centroid. For lag $\tau$, the corresponding
mean-squared displacement (MSD) satisfies
\begin{align}
    M_i(\tau)
    &=
    \left\langle
    \left|
    \Delta_\tau\bm x_i(t)
    \right|^2
    \right\rangle
    \nonumber\\
    &=
    \underbrace{
    \left\langle
    \left|
    \Delta_\tau\bm X_{\mathrm c}(t)
    \right|^2
    \right\rangle
    }_{M_{\mathrm c}(\tau)}
    +
    \underbrace{
    \left\langle
    \left|
    \Delta_\tau\bm r_i(t)
    \right|^2
    \right\rangle
    }_{M_{\mathrm rel}(\tau)}
    +
    2
    \underbrace{
    \left\langle
    \Delta_\tau\bm X_{\mathrm c}(t)
    \cdot
    \Delta_\tau\bm r_i(t)
    \right\rangle
    }_{C_{\mathrm{c,rel}}(\tau)}.
    \label{eq:msd_decomposition}
\end{align}
This separates displacement caused by translation of the whole team from
rearrangement within the moving formation.

\begin{figure}[!htbp]
    \centering
    \FullWidthFigure{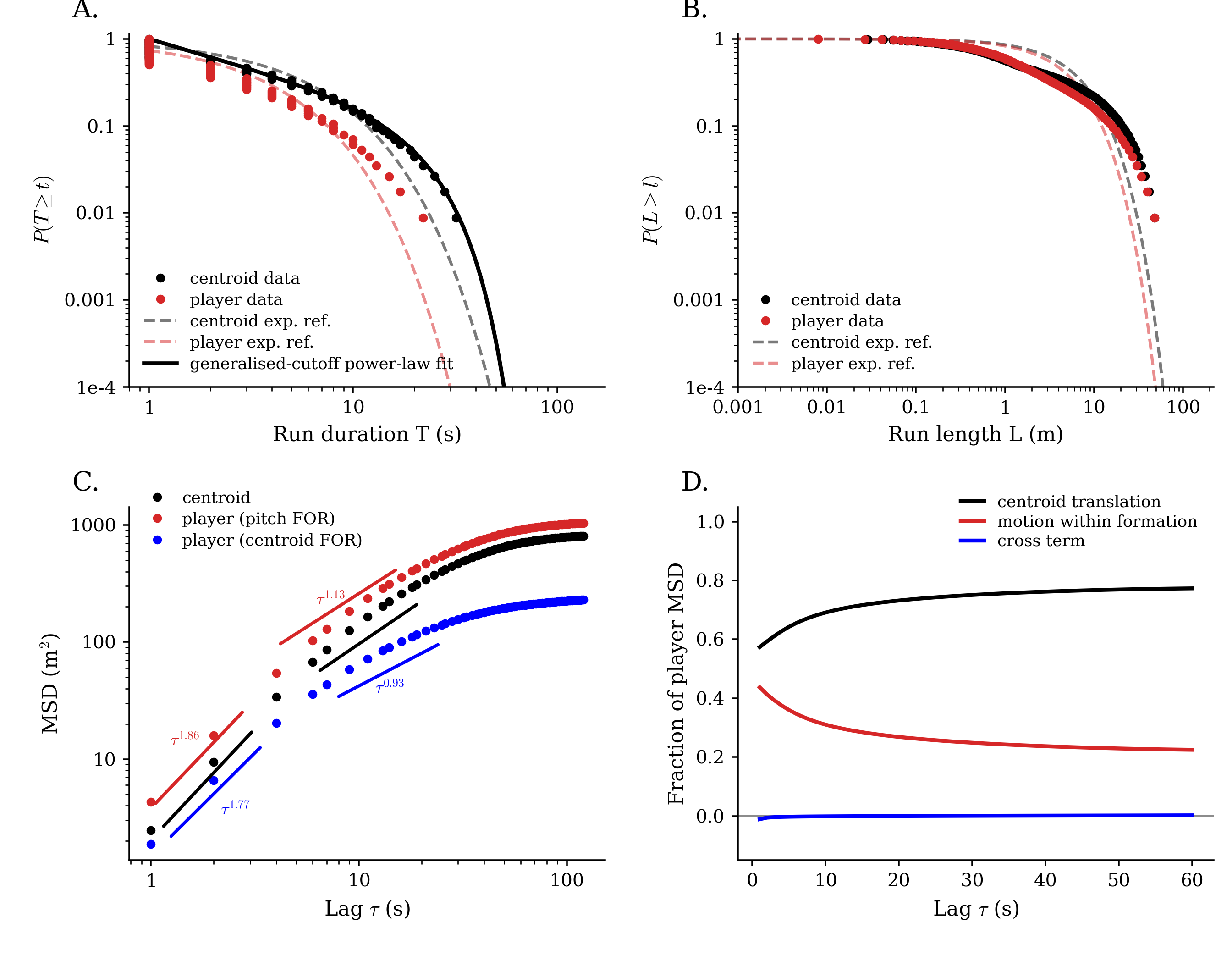}
    \caption{
    \textbf{Player transport is increasingly carried by collective translation.}
    \textbf{A}, Empirical complementary cumulative distributions of centroid-
    and player-run duration. Dashed curves are exponential references matched
    separately to the corresponding empirical means; the solid black curve is
    the best-supported generalised-cutoff power-law fit to completed centroid
    runs. Formal model comparison is reported in Supplementary Fig.~S1 and
    Supplementary Table~S2.
    \textbf{B}, Run-length distributions and descriptive mean-matched
    exponential references.
    \textbf{C}, MSD of the team centroid, players in the pitch frame and players
    in the team-centroid-relative frame. Short-lag guide slopes over 1--4~s
    illustrate a strongly superdiffusive, near-ballistic persistence regime;
    the principal effective exponents were fitted over 5--30~s to quantify the
    subsequent frame-dependent crossover.
    \textbf{D}, Exact signed decomposition of pitch-frame player MSD into
    centroid, relative-motion and centroid--relative covariance contributions.
    Uncertainty in Figure~2 was estimated using 200 cluster-bootstrap
    replicates over continuous trajectory sources, clustered by match, team and
    match phase.
    }
    \label{fig:transport}
\end{figure}

Directional-run durations were broader than mean-matched exponential
distributions for both centroid and player trajectories
(\FigRef{fig:transport}A). These exponentials are memoryless references: after
matching the empirical mean, they represent a constant event rate that does
not depend on how long a run has already persisted. Centroid runs had higher
survival probability at intermediate and long durations, indicating greater
directional persistence at the collective scale.

Across 69,670 completed centroid runs, we compared six candidate duration
families using grouped leave-one-match-date-out log likelihood. The
generalised-cutoff model---a power law with a stretched-exponential cutoff, in
which $\beta=1$ recovers conventional exponential truncation---ranked first,
ahead of the conventional truncated power law, Weibull, log-normal, pure power
law and geometric model, the discrete-time analogue of a memoryless
exponential (Supplementary Fig.~S1 and Supplementary Table~S2). The
whole-sample fit yielded $\gamma=1.33$, $T_c=32.3~\mathrm{s}$ and
$\beta=2.78$. Here, $\gamma$ controls the broad early-time decay, $T_c$ sets
the characteristic cutoff timescale and $\beta$ controls the sharpness of the
late attenuation. Because $\beta>1$, the tail was attenuated more sharply than
under conventional exponential truncation. None of the candidate families
reproduced the full empirical distribution closely enough to pass the absolute
goodness-of-fit test. The generalised-cutoff model is therefore used as the
best relative summary among the families tested, not as an exact generative
law.

This fit describes completed run lifetimes; the interval-level termination
process is analysed below. Run-length survivors were likewise broader than
their mean-matched exponential references (\FigRef{fig:transport}B), although
we did not assign a unique distributional family to length. The late downturn
in both duration and length is consistent with finite pitch and match
boundaries and inconsistent with an unbounded pure power-law tail over the
observed range.

The MSD curves showed an initial persistence regime followed by a
frame-dependent crossover (\FigRef{fig:transport}C). If
$M(\tau)\propto\tau^{\alpha}$, ordinary diffusion corresponds to $\alpha=1$,
ballistic motion to $\alpha=2$, and $1<\alpha<2$ indicates superdiffusion. At
short lags of 1--4~s, all three coordinate systems were strongly
superdiffusive, with descriptive slopes close to the ballistic limit. Over the
principal 5--30~s fitting range, the dynamics separated: centroid and
pitch-frame player MSDs remained superdiffusive, with effective exponents of
approximately $1.20$ and $1.13$, whereas centroid-relative motion became
approximately diffusive, with an exponent of $0.93$. Thus, after the first few
seconds, directional persistence remained in the centroid and pitch-frame
coordinates but was largely lost within the formation. These slopes describe
a finite-range crossover rather than asymptotic transport.

The MSD decomposition showed more directly how player displacement was divided
between team translation and internal motion (\FigRef{fig:transport}D).
The centroid contribution increased with lag, the relative-motion contribution
decreased, and the centroid--relative cross term remained small. At longer
timescales, translation of the team as a whole accounted for an increasing
share of absolute player displacement, while the contribution from
rearrangement within the formation declined.

The player- and centroid-level signatures were therefore coupled rather than
independent. Individual trajectories still contained motion not captured by
the centroid, but at longer lags a growing share of player displacement came
from translation of the team. The centroid thus provides a natural collective
coordinate for studying how directional team movements persist and
reorganise.

\FloatBarrier

\subsection*{Collective order during run onset is associated with persistence}

We next tested whether collective order near run onset was associated with how
long a run persisted. Order was quantified by the polarisation
\begin{equation}
    p(t)
    =
    \left|
    \frac{1}{N_v(t)}
    \sum_{i\in\mathcal A_v(t)}
    \frac{\bm v_i(t)}
    {|\bm v_i(t)|}
    \right|,
    \qquad
    0\leq p\leq1,
    \label{eq:polarisation}
\end{equation}
where $\mathcal A_v(t)$ contains active players with a defined non-zero
velocity direction and $N_v(t)=|\mathcal A_v(t)|$. Thus,
$\mathcal A_v(t)\subseteq\mathcal A(t)$: all selected active players with
valid coordinates contribute to the centroid, whereas only those with a
defined non-zero velocity direction contribute to polarisation. Values near one
indicate aligned motion, whereas lower values indicate less coherent movement.
For the descriptive Figure~3 analysis, low,
mid and high denote global terciles of run-onset order across the pooled runs,
rather than universal absolute levels of polarisation. The cross-fitted
interval analyses in Figures~4 and 5 instead used team-specific cutpoints
estimated from the training match dates and applied unchanged to the excluded
date group.

\begin{figure}[!htbp]
    \centering
    \FullWidthFigure{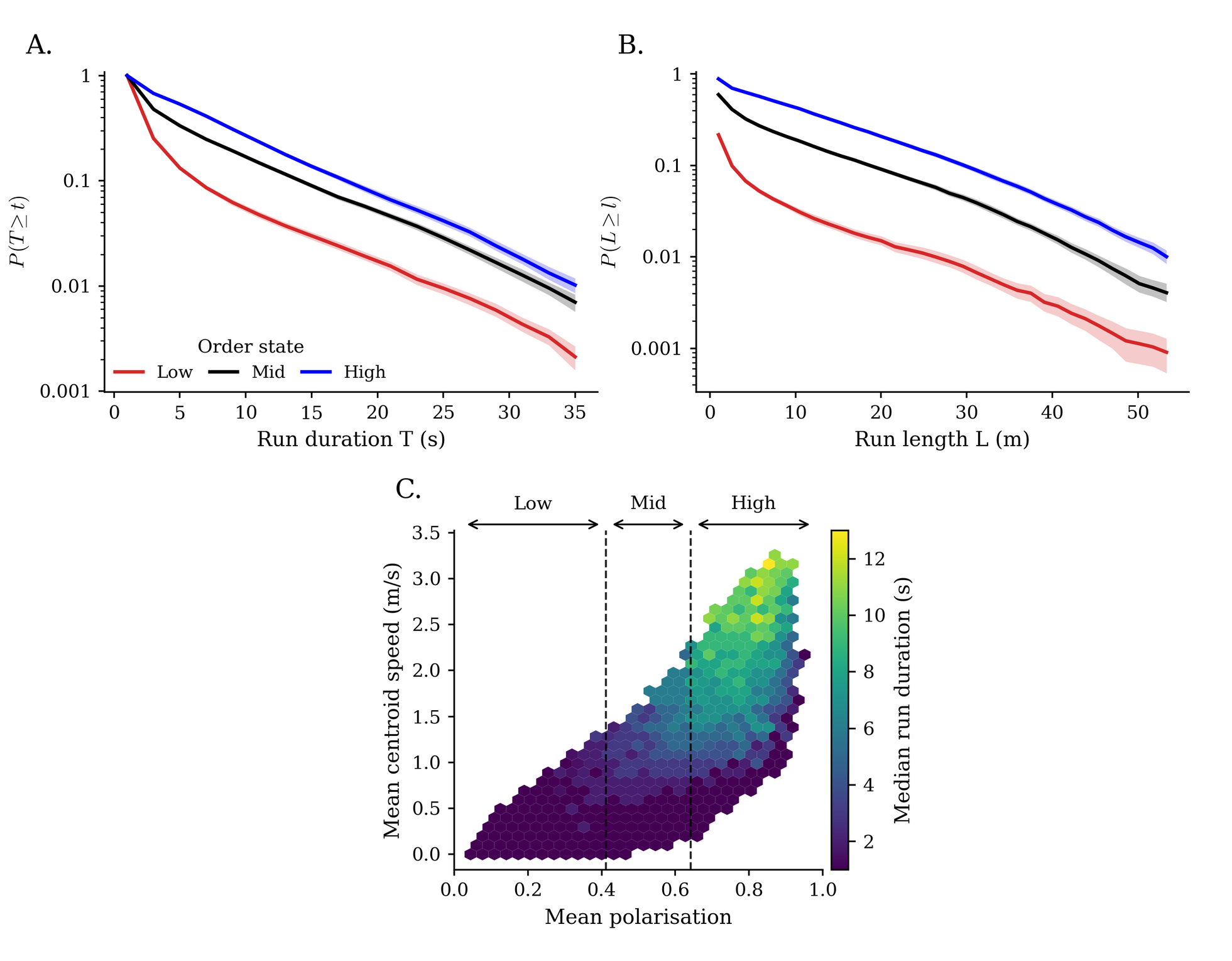}
    \caption{
    \textbf{Collective order during run onset is associated with persistence.}
    \textbf{A}, Duration distributions for runs classified by global low,
    mid or high terciles of mean polarisation over the first up to 3 s; runs
    shorter than 3 s use all available samples.
    \textbf{B}, Run-length distributions for the same onset-order classes.
    Ribbons in \textbf{A--B} show 95\% cluster-bootstrap intervals from 500
    resamples of season--match-date--team groups.
    \textbf{C}, Joint distribution of whole-run mean polarisation and mean
    centroid speed; colour denotes median run duration. This panel is a
    descriptive state-space summary, and the coupling between order and speed
    motivates the kinematic adjustments in the interval-level analyses.
    }
    \label{fig:order}
\end{figure}

For descriptive run-level stratification, onset order was the mean
polarisation over the first up to 3 s of each run. Runs shorter than 3 s were
retained and averaged over all available samples. High-order runs had the
broadest duration distribution, low-order runs the narrowest, and mid-order
runs lay between them (\FigRef{fig:order}A). Path lengths followed the same
ordering (\FigRef{fig:order}B). For runs shorter than 3 s, the onset window
necessarily included the endpoint; Figure~3 therefore describes an association
rather than a prospective prediction.

Order and speed were closely related. High-polarisation runs occurred more
often at higher centroid speeds, and long runs occupied an ordered,
faster-moving region of the joint state space (\FigRef{fig:order}C). This
relationship creates a potential kinematic confound: faster translation can
increase run length without extending duration, while tactical context may
influence both speed and order.

We then tested order--termination associations at the interval level. The
primary speed-adjusted analysis used current interval-level polarisation, which
is contemporaneous with the one-second interval outcome rather than strictly
preceding it. After adjustment for interval age and mean active-player speed, a
one-standard-deviation increase in current polarisation was associated with a
termination odds ratio of $0.80$ (95\% match-date bootstrap interval
$0.79$--$0.81$). Thus, conditional on age and mean active-player speed, the
odds of termination were approximately 20\% lower per standard-deviation
increase in polarisation. The estimate remained below one within
training-defined slow, middle and fast speed strata and after velocity
coherence was removed from active-player selection (Supplementary Table~S2).

A stricter analysis averaged polarisation over the preceding five seconds,
excluding the current interval. This lagged measure remained inversely
associated with termination after separate adjustment for lagged mean
active-player speed (odds ratio $0.82$, 95\% interval $0.80$--$0.83$) or lagged
centroid speed ($0.94$, $0.93$--$0.96$). Including both speed measures moved
the estimate close to no association, for which the odds ratio is one
($0.98$, $0.96$--$1.00$). The two speed measures were strongly correlated
($r=0.92$), and a maximum variance-inflation factor of $8.0$ indicated
substantial collinearity; the joint model was therefore treated as a diagnostic
rather than the primary adjustment. These estimates quantify associations and
do not establish a causal effect of collective order.

Adding preceding-five-second polarisation to models already containing run age
and preceding centroid speed produced a small improvement in held-out log loss
on excluded match dates. The improvement remained after including preceding
centroid directional stability, measured from the consistency of centroid
headings over the same five-second history. The corresponding Brier-score
contrasts included zero, and the gains varied across match dates. Polarisation
therefore added only limited predictive information beyond centroid
kinematics (Supplementary Table~S2).

\FloatBarrier

\subsection*{Age and preceding order predict survival on held-out match dates}

\begin{figure}[!htbp]
    \centering
    \FullWidthFigure{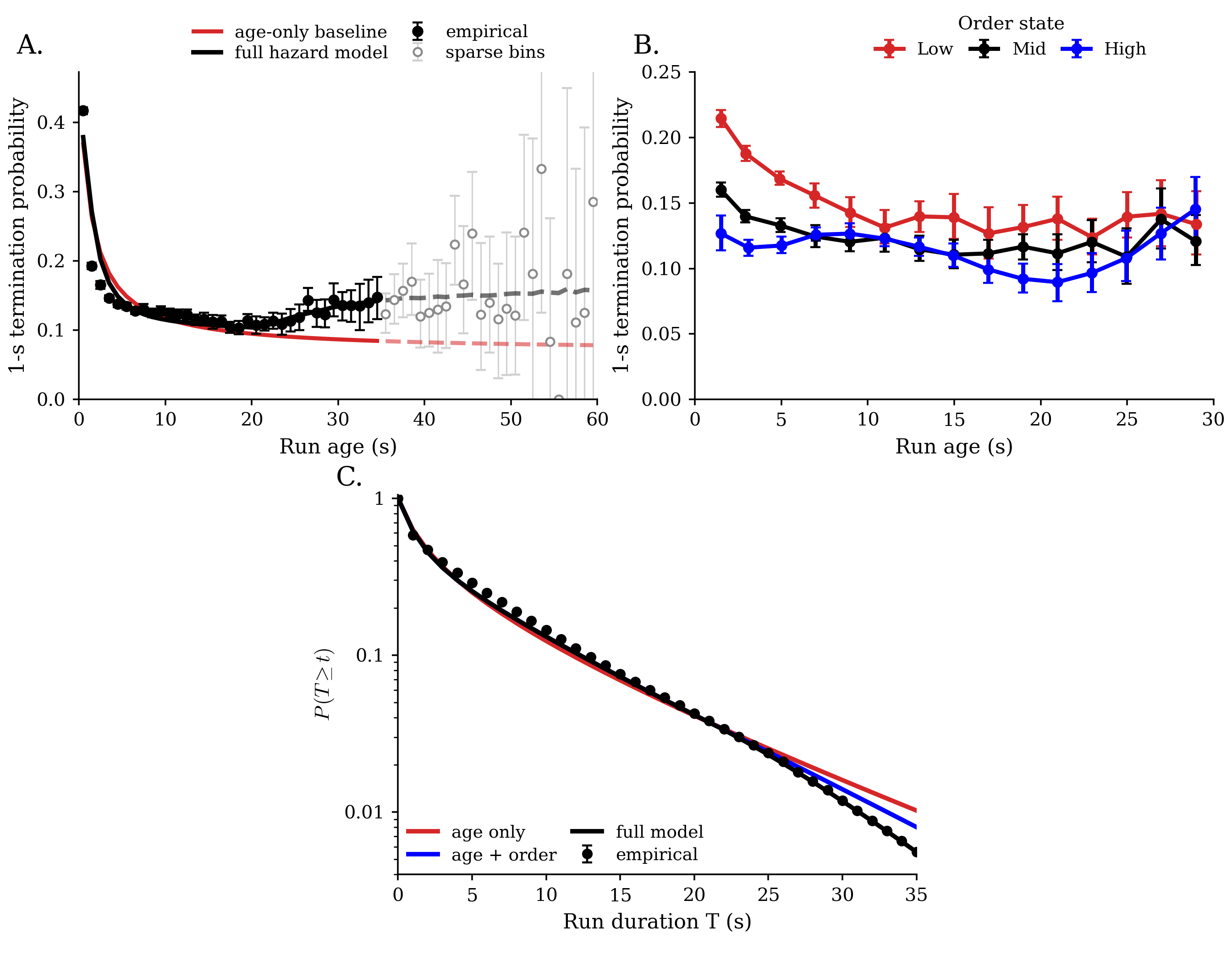}
    \caption{
    \textbf{Age- and state-dependent termination predicts run survival.}
    \textbf{A}, Pooled empirical one-second termination probability versus run
    age, together with the pooled age-only baseline and the model containing
    age, preceding order and the later-age term. The vertical
    line at 35 s marks the upper boundary of the primary supported range.
    Model continuation and sparse empirical bins beyond that boundary are shown
    for context only; open grey markers denote sparsely supported bins.
    \textbf{B}, Pooled empirical termination probability stratified by the modal
    low-, mid- or high-order state among available intervals in the strictly
    preceding 5 s; the current interval is excluded. Error bars in \textbf{A--B}
    show 95\% intervals from 500 match-date bootstrap replicates.
    \textbf{C}, Grouped leave-one-match-date-out observed-path survival
    prediction. At each fold, hazard parameters and state thresholds were
    estimated from all other match dates. Order-dependent models used the
    observed held-out order sequence only as a time-varying covariate; no
    independent uncertainty band is shown. The age-only model contains the
    inverse-age component, age plus order adds preceding-state multipliers, and
    the full model additionally includes the gradual later-age term.
    }
    \label{fig:termination}
\end{figure}

We next examined how termination changed with run age and whether preceding
order improved survival prediction on held-out match dates. Let $\mathcal T$
denote a run lifetime. Its survival function and continuous-time termination
hazard satisfy
\begin{equation}
    S(a)
    =
    \Prob(\mathcal T>a),
    \qquad
    \lambda(a)
    =
    -
    \frac{\dd}{\dd a}
    \log S(a).
    \label{eq:hazard_identity}
\end{equation}
The empirical analysis was performed on one-second intervals, so
\FigRef{fig:termination}A--B displays interval termination probability rather
than the infinitesimal hazard itself.

Termination probability was highest immediately after a run began and then
declined sharply
(\FigRef{fig:termination}A).
Over the early and intermediate supported range, this decrease was approximated
by an inverse-age baseline,
\begin{equation}
    \lambda_0(a)
    =
    \frac{\mu}{a+a_0},
    \label{eq:inverse_age_hazard}
\end{equation}
which produces a power-law form of survival when acting alone. Empirical
termination reached a minimum around the middle of the supported range and
increased modestly thereafter.

We represented the crossover with the effective state-dependent hazard
\begin{equation}
    \lambda_k(a)
    =
    m_{s_k(a)}
    \frac{\mu}{a+a_0}
    +
    q
    \left[
    1-
    \exp
    \left(
    -
    \frac{
    \positivepart{a-t_c}
    }{
    \tau_q
    }
    \right)
    \right],
    \label{eq:full_hazard}
\end{equation}
where $s_k(a)$ is the recent low-, mid- or high-order state of run $k$,
$m_{s_k(a)}\in\{m_L,m_M,m_H\}$ is the multiplier selected by that state, and
$(x)_+=\max(x,0)$. The state multiplier applies only to the inverse-age
component; the later-age contribution is common to all three states. The second
term is zero before $t_c$ and rises gradually thereafter, with $t_c$, $\tau_q$
and $q$ determining its onset, rise time and magnitude. This later-age term
describes the late attenuation of survival over the observed range; the
available data do not resolve the hazard at much longer ages.

At a given age, termination also varied with recent order
(\FigRef{fig:termination}B).
Low-order intervals were generally most vulnerable, high-order intervals least
vulnerable, and mid-order intervals intermediate over the well-supported range.
For the cross-fitted age-plus-order model, the median state multipliers were
$1.31$, $1.00$ and $0.67$ for low, mid and high order, respectively.
Thus, within the inverse-age component of the fitted model, high-order intervals
carried approximately half the termination multiplier of low-order intervals.
The later-age contribution began at approximately 20 s and rose over a
timescale of approximately 5 s. Exact fold-median parameters are reported in
Supplementary Methods S4. The state separation weakened and uncertainty
increased towards the oldest supported ages.

Generalisation was evaluated in 47 folds. In each fold, all tracked records
from one match date were held out and the models were fitted to the remaining
46 date groups. Fifteen folds contained two separate fixtures played by the
tracked clubs on the same date; the others contained one fixture
(\FigRef{fig:termination}C).

The models formed a nested sequence: age only; age plus the strictly preceding
low-, mid- or high-order state; and age plus order plus the gradual later-age
term in Eq.~\eqref{eq:full_hazard}. The last model allows termination
probability to rise again after approximately 20 s.

These were observed-path predictions. The fitted parameters came from the
training dates, while the held-out preceding-order sequence was supplied as a
time-varying covariate. Figure~4C therefore tests transfer of the termination
relationship rather than prediction of future order. The age-only model
predicted too many long-lived runs, adding order reduced this discrepancy, and
the later-age term captured the attenuation most closely.

Held-out scores confirmed the same ordering. Log loss decreased from $0.471$
for the age-only model to $0.466$ with order and $0.464$ for the full model.
The full model was also close to ideal calibration, with slope $0.99$ and
intercept $-0.03$. The Brier score changed only slightly, from $0.148$ to
$0.147$, indicating that the clearest gain was improved likelihood-based fit
and calibration of interval-level termination probabilities rather than a
large increase in overall predictive accuracy.

\begin{table}[!htbp]
\centering
\caption{\textbf{Held-out interval-level termination prediction.}
All values were calculated from pooled out-of-fold predictions for match-date
groups excluded from model fitting. Lower log loss and Brier score indicate
better probability predictions; ideal calibration has slope 1 and intercept 0.}
\small
\begin{tabular}{lrrrr}
\toprule
Model
& Log loss $\downarrow$
& Brier score $\downarrow$
& Calibration slope
& Calibration intercept\\
\midrule
Age only
& $0.471$
& $0.148$
& $0.81$
& $-0.24$\\
Age + order
& $0.466$
& $0.148$
& $0.90$
& $-0.12$\\
Age + order + later-age term
& $0.464$
& $0.147$
& $0.99$
& $-0.03$\\
\bottomrule
\end{tabular}
\label{tab:hazard_validation}
\end{table}

Held-out evaluation included 355,330 one-second intervals and 69,670 observed
terminations. The remaining 132 run endings were right-censored; they
contributed exposure through the final observed interval but were not counted
as termination events.

The full model continued to underestimate termination over part of the
early-age range, but both preceding order and the later-age term improved
prediction on excluded match dates.

The age-dependent pattern also persisted under alternative segmentation rules
(Supplementary Fig.~S2 and Supplementary Table~S2). A centred three-sample rolling median of the one-second coordinates produced
nearly the same run count, duration scale and termination curve as the primary
segmentation. In the low-speed guard analyses, an apparent turn was suppressed
when either adjacent centroid step was slower than $0.25$ or
$0.50~\mathrm{m\,s^{-1}}$. Changing the treatment of low-speed turns produced fewer, longer runs and
shifted the absolute termination probabilities, but the onset peak, decline
and later flattening remained.

\FloatBarrier

\subsection*{Age-dependent switching predicts the composition of surviving runs}

We next asked whether state switching and differential termination could
account for the changing order composition of surviving runs. Figure~4
conditions termination on a modal state derived from the strictly preceding
5 s; here we follow the one-second evolution of the current collective state.
For each fold, current polarisation was divided into team-specific low, mid and
high terciles using only the training match dates. Let
$Z_j\in\{L,M,H\}$ denote this current state in interval $j$. For age band $b(j)$, the
transition probability conditional on survival is
\begin{equation}
    P_{sr,j}
    =
    \Prob\!\left(
    Z_{j+1}=r
    \mid
    Z_j=s,\;\text{run survives interval }j
    \right),
    \qquad
    \bm P_j=\bm P^{(b(j))},
    \label{eq:conditional_transition}
\end{equation}
where $b(j)$ is one of $0$--$5$, $5$--$10$, $10$--$20$ or $20$--$35$ s.
Switching and state-specific termination were indexed by this same
fold-thresholded current-state definition.

\begin{figure}[!htbp]
    \centering
    \FullWidthFigure{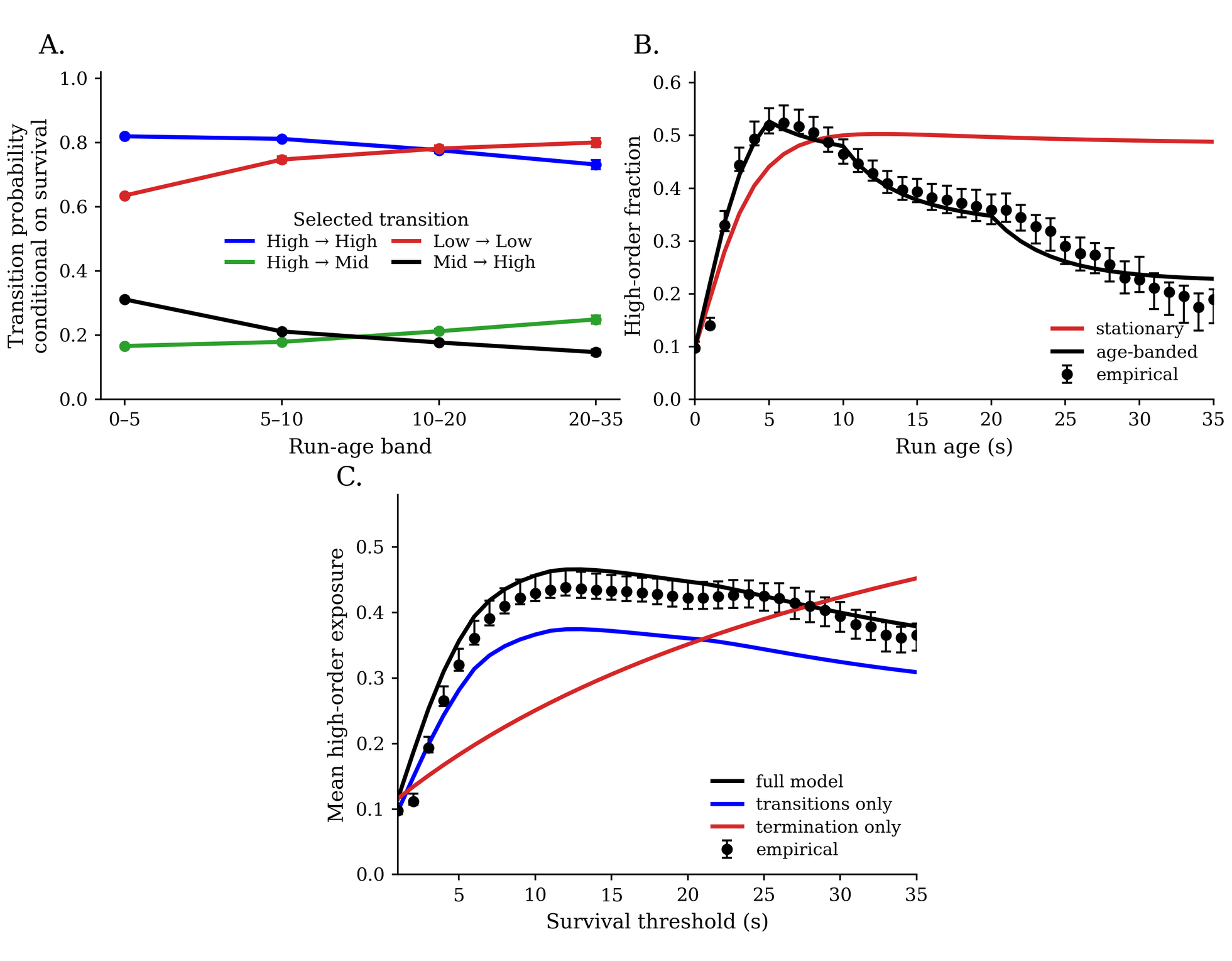}
    \caption{
    \textbf{Age-dependent switching and differential termination shape the
    composition of long-lived runs.}
    \textbf{A}, Pooled selected one-second transition probabilities between
    current order states across age bands, conditional on survival of the
    interval. Complete $3\times3$ matrices are reported in the Supplement.
    \textbf{B}, Grouped leave-one-match-date-out prediction of the current
    high-order fraction among surviving runs. The stationary transition model,
    which uses the same transition matrix at every age, misses the empirical
    trajectory, whereas the age-banded model reproduces its rise and
    subsequent decline. Predictions use only
    training-estimated birth distributions, age-dependent transition matrices
    and state-dependent termination probabilities.
    \textbf{C}, Held-out mean high-order exposure, defined as the fraction of
    elapsed run time spent in the high-order state, compared with the full
    switching-and-termination model and two simplified process-removal models. The
    switching-with-common-termination model retains age-banded switching but
    removes differences in termination probability between order states. The
    termination-without-switching model retains state-dependent termination but
    prevents switching by setting each transition matrix to the identity. Because each modified model changes the population entering later updates,
    the two comparisons should not be interpreted as additive effects.
    Error bars show 95\% intervals from 500 match-date bootstrap replicates.
    }
    \label{fig:state_model}
\end{figure}

Transition probabilities changed systematically with age
(\FigRef{fig:state_model}A). High-order persistence decreased from
approximately $0.82$ in the first age band to $0.73$ in the last, while
low-order persistence increased from approximately $0.64$ to $0.80$.
Mid-to-high switching decreased from approximately $0.31$ to $0.15$, whereas
high-to-mid switching increased from approximately $0.17$ to $0.25$. Direct
low-to-high transitions were uncommon; early movement towards high order
occurred mainly through low-to-mid and mid-to-high steps.

To propagate the whole surviving population, let
\[
\bm u_j
=
\begin{pmatrix}
 u_{L,j}\\
 u_{M,j}\\
 u_{H,j}
\end{pmatrix}
\]
record the fraction of the original run cohort that remains alive in each state
at interval $j$. With $d_{s,j}$ the probability that a run in current state
$s$ terminates during interval $j$, the fitted population model evolves as
\begin{equation}
    \bm u_{j+1}
    =
    \bm A_j^{\mathsf T}\bm u_j,
    \qquad
    A_{sr,j}
    =
    \left(1-d_{s,j}\right)P_{sr,j}.
    \label{eq:discrete_state_model}
\end{equation}
The matrix $\bm A_j$ combines survival during the current interval with
switching into the next state. Rows of $\bm P_j$ denote the current state and
columns the next state; the transpose appears because $\bm u_j$ is written as
a column vector. Total survival and survivor composition are
\begin{equation}
    S_j=\bm 1^{\mathsf T}\bm u_j,
    \qquad
    c_{s,j}=\frac{u_{s,j}}{S_j}.
    \label{eq:survival_composition}
\end{equation}
Here, $\bm 1$ is a vector of ones, so $S_j$ is obtained by summing the three
live-state components. Once the distribution of order states at run birth
(the birth-state distribution), transition matrices and termination
probabilities have been estimated, future state composition is generated
without replaying an observed held-out state sequence.

Current order and cumulative exposure to a state capture different aspects of
a run. For runs surviving to age threshold $A>0$, mean exposure to state $s$
was
\begin{equation}
    \Pi_s(A)
    =
    \Expect\!\left[
    \frac{1}{A}
    \int_0^A
    \mathbf 1\!\left\{Z(a)=s\right\}
    \dd a
    \;\middle|\;
    \mathcal T\geq A
    \right],
    \label{eq:state_exposure}
\end{equation}
where $Z(a)$ is the categorical current-state process in continuous-age
notation.

The empirical high-order fraction among surviving runs rose from approximately
$0.1$ at birth to above $0.5$ around 5--7 s and then declined towards $0.2$ by
35 s (\FigRef{fig:state_model}B). A stationary comparator, which used the same transition matrix at every run
age, approached a nearly constant high-order fraction and failed to reproduce
this trajectory. The age-banded model, fitted without each evaluated match-date
group, generated the early rise, peak and later decline. The model underestimated the high-order fraction at some intermediate ages
and slightly overestimated it in the oldest bins, but reproduced the main
rise-and-decline pattern without using future states from the held-out matches.

The same switching-and-termination process can also be written in continuous age. Let
$\bm n(a,t)$ denote the density of live runs of age $a$, match time $t$ and
current order state. Let $\bm K(a)$ be a continuous-time switching generator
whose off-diagonal element $\kappa_{sr}(a)$ is the transition rate from state
$s$ to state $r$ and whose rows sum to zero. With
\[
\bm D(a)
=
\operatorname{diag}\!\left[
\lambda_L(a),\lambda_M(a),\lambda_H(a)
\right],
\]
the formal continuous-age limit is
\begin{equation}
    \frac{\partial\bm n}{\partial t}
    +
    \frac{\partial\bm n}{\partial a}
    =
    \left[\bm K(a)^{\mathsf T}-\bm D(a)\right]\bm n.
    \label{eq:mckendrick_state}
\end{equation}
This is a McKendrick--von Foerster interpretation of the fitted discrete model,
not a separately fitted smooth generator
\cite{McKendrick1926,VonFoerster1959}. The continuum form places the recursion
within the standard class of age-structured population-balance models and makes
explicit the distinction between mass-conserving redistribution by switching
and removal of live runs by termination.

For a birth cohort, define $S(a)=\bm 1^{\mathsf T}\bm u(a)$,
$\bm c(a)=\bm u(a)/S(a)$ and
$\overline{\lambda}(a)=\sum_s c_s(a)\lambda_s(a)$. Because the transition
generator conserves live mass,
\begin{equation}
    \frac{\dd S}{\dd a}
    =
    -\overline{\lambda}(a)S(a).
    \label{eq:total_survival_balance}
\end{equation}
Switching redistributes live runs among states but does not terminate them
directly. Survivor composition therefore obeys
\begin{equation}
    \frac{\dd\bm c}{\dd a}
    =
    \underbrace{\bm K(a)^{\mathsf T}\bm c(a)}_{
    \text{redistribution by state switching}}
    +
    \underbrace{\left[\overline{\lambda}(a)\bm I-\bm D(a)\right]\bm c(a)}_{
    \text{selection by differential termination}}.
    \label{eq:composition_balance}
\end{equation}

The full switching-and-termination model reproduced the rapid early increase,
the extended peak and the later decline in high-order exposure (\FigRef{fig:state_model}C). To separate
the roles of the two processes, we reran the model in two simplified forms.
When differences in termination between states were removed, age-dependent
switching still generated the rapid early rise in high-order exposure. When
switching was prevented, state-dependent termination produced a slower
enrichment of high-order runs. The full switching-and-termination model was required to reproduce both the
early rise and its persistence at later ages. Because each process changes the
population on which the other subsequently acts, their effects are not
additive.

\FloatBarrier

\section*{Discussion}

Our results connect player displacement, collective order and the lifetime of
team-centroid runs. At longer lags, player motion increasingly reflected
translation of the team. These translating runs were most vulnerable soon
after onset; termination then declined with age, varied modestly with recent
order and rose again later. Because runs also switched between order states,
the composition of long-lived runs reflected both reorganisation and selective
termination. Broad collective transport therefore emerged from the histories
of modes that persisted and changed state, rather than from durations fixed
when a run began.

Previous football studies have largely treated spatial organisation and
transport as separate problems. Team centroids, occupied areas, pairwise
coordination and recurrent collective states describe how teams are organised
\cite{Frencken2011,Bartlett2012,Moura2013,Welch2021,Marcelino2020}, while
stochastic models have examined how player interactions generate team-level
dynamics \cite{Chacoma2021}. More recently, broad run statistics in players and
team centroids motivated the suggestion that football movement resembles
collective foraging \cite{Shpurov2024}. Our results link these strands: the
state of a translating centroid run was associated with its persistence, and
age-dependent transitions determined which forms of organisation remained
among older runs. Collective state therefore describes not only how a team is
organised, but also how long its collective movement persists.

The importance of the centroid depended on timescale. At the shortest lags,
motion in all reference frames was strongly persistent. Over the subsequent
regime, motion within the formation lost most of its directional memory,
whereas centroid and pitch-frame motion remained superdiffusive. The centroid
is therefore not merely a spatial average; it captures the slower translation
of the team that continues to shape player displacement after internal motion
has become nearly diffusive. This separation of timescales explains why player-
and team-level transport signatures can coexist without being independent. At
short times the player remains the relevant object; at longer times the
translating team mode carries an increasing fraction of the motion.

These findings point to a route to L\'evy-like transport that does not require
an explicit search programme. Heavy-tailed movement statistics do not by
themselves identify a search strategy or a unique stochastic process.
Interactions can generate history-dependent, or non-Markovian, persistence
\cite{FedotovKorabel2017}; swarm fluctuations can create broad transport
without a prescribed L\'evy programme \cite{ReynoldsOuellette2016}; and
age-dependent turning or mortality can temper otherwise persistent walks
\cite{FedotovTanZubarev2015,Stage2017}. Here, the broad finite-range
distribution arises from collective modes whose switching and termination
rates change with age and state. The generalised-cutoff distribution is
therefore an empirical summary of the resulting lifetimes, not evidence that a
team samples a duration when a run begins.

Age was the dominant predictor of termination. Mild rolling-median smoothing
left the primary termination curve and duration scale almost unchanged, while
low-speed turn guards produced fewer and longer runs but preserved the high
onset risk and subsequent decline. The exact timescale and later-age shape
depended on how low-speed directional changes were treated, so the late rise
should be interpreted as finite-range tempering rather than a universal
asymptotic law.

The association with order was consistent but modest. Lagged polarisation
added a small amount of held-out predictive information beyond centroid speed
and recent directional stability, but much of its unadjusted association was
shared with collective kinematics. Polarisation should therefore be viewed as
one feature of a persistent translating state, not as a large independent
protective force. The available data cannot distinguish whether alignment
helps stabilise translation, records an already stable mode or lies on the same
causal pathway as directional persistence.

The state model extends classical survivor selection to a population whose
risk state changes over time. In classical frailty models, aggregate mortality can decline
because vulnerable individuals are removed earlier from a heterogeneous
population \cite{Vaupel1979}. Here, heterogeneity was partly observed through
collective order and was not fixed at birth. Runs switched between low-, mid-
and high-order states while differential termination continually altered the
population that remained. Long-lived runs were therefore not simply a
pre-existing class of unusually persistent movements. The process-removal
models showed that switching generated the early increase in high-order
exposure, whereas differential termination helped sustain that enrichment.
Together, the two processes explain why the order composition of older runs
differs from that of newly formed runs.

These results do not exclude foraging as a functional description of football
movement. Search for the ball, open space or advantageous configurations may
influence when collective runs begin, how states switch or when runs end. Our
analysis addresses a different question: how those contextual influences
become broad trajectory statistics. With richer event data, the foraging
hypothesis could be tested more directly by asking whether possession, ball
distance, opponent pressure or access to space modifies run formation,
state-transition probabilities or state-dependent termination. Foraging and
survival dynamics may therefore operate at different levels of explanation:
one concerns why movement occurs, the other how its statistical form is
generated.

The same framework could be applied to other interacting groups by identifying
an appropriate collective coordinate, segmenting its finite-lived movements
and estimating how age and internal state affect termination. Candidate
systems include animal groups, migrating cell collectives, pedestrian crowds,
robot swarms and other team sports.

Several limitations constrain the present interpretation. The cohort contains
two clubs over two seasons, and match-date cross-fitting provides internal
rather than external validation. The state definitions and age bands emerged
during exploratory analysis, and run counts and duration scales remain
sensitive to segmentation. Active players were reconstructed without complete
official substitution records, while incomplete ball, possession, opponent
and event data prevent identification of the football event that ends each
run. These limitations restrict causal and tactical interpretation and leave
the generality of the proposed mechanism to be tested in other teams,
competitions and tracking systems.

A trajectory records selection among histories as well as the rules of motion.
A completed long movement need not have been chosen as long. It may be the
realised path of a collective mode that avoided termination, changed state and
remained coherent long enough to travel far. In this system, player motion
increasingly projected onto team translation, the lifetime of that translation
depended on age and collective state, and the surviving population was
reshaped by switching and differential termination. The object requiring
explanation is therefore the collective mode that survives long enough to
produce the long step.

\section*{Methods}

\subsection*{Tracking data and preprocessing}

We analysed publicly available SoccerMon position records for Rosenborg and
V\aa lerenga during the 2020 and 2021 Norwegian women's top-flight Toppserien
seasons \cite{Midoglu2024,SoccerMonZenodo}. Official Norwegian Football
Federation schedules and results were used to identify competitive fixtures
and exclude non-match recordings on the same dates
\cite{NFFToppserien2020,NFFToppserien2021}. We analysed two fixed match-play
windows: scheduled kickoff to $+45$ min and $+60$ to $+105$ min.

The cohort comprised 66 tracked team-match records from 62 fixtures played on
47 match dates. Four head-to-head fixtures contributed one record for each
club. On 15 other dates, the two clubs played separate opponents; these
fixtures were grouped together only for date-level validation. The final
analysis contained 69,802 centroid runs, including 69,670 observed directional
terminations and 132 right-censored endings, and 355,330 one-second intervals
at risk.

Venue-specific pitch geometry was obtained from OpenStreetMap through the
Overpass API, and latitude and longitude were converted to pitch-aligned
Cartesian coordinates \cite{OpenStreetMap}. Native positions were recorded at
10 Hz with STATSports APEX multi-GNSS units \cite{Midoglu2024}. To align the
trajectories with the one-second termination and switching models, and to keep
the full multiseason analysis computationally tractable, records were reduced
to a common 1-Hz time base by retaining the first timestamp-ordered observation
in each second. Missing seconds were left missing, so interpolated positions
could not create turns, velocities or state transitions. The primary analysis
used unsmoothed coordinates; smoothing was examined only as a segmentation
sensitivity.

After common preprocessing, records were pooled at the level relevant to each
analysis. Duration and length distributions used all runs, empirical
termination probabilities pooled events and risk sets within each age bin, and
descriptive transition matrices pooled one-second transition counts. These
observations were not treated as independent: bootstrap resampling preserved
match-date or match--team--phase--source grouping. Cross-fitted outputs were
pooled only after each interval or state trajectory had received a prediction
from a model that excluded its match date. Supplementary Methods S2 and S5
provide additional processing and pooling details.

\subsection*{Active-player reconstruction}

Team recordings included all instrumented players, including substitutes and
off-pitch tracks. We therefore reconstructed the on-field set at every
one-second frame before calculating the centroid or collective order. A
candidate required finite coordinates within a 30-m coordinate-sanity margin
and, unless selected in the preceding frame, could lie no more than 5 m outside
the calibrated pitch.

Pitch location also contributed a persistent participation state. Remaining at
least 6 m inside the pitch for 70 s supported an active geometry state, whereas
110 s off the pitch supported demotion from that state. This geometry state
was one component of the final selector rather than the active-player label
itself. The initial set comprised the 11 highest-scoring players over the first
120 s. Thereafter, candidates were ranked by recent movement, pitch position,
geometry state, initial-set membership, directional coherence and continuity.
Vacancies were filled immediately; when 11 players were already selected,
replacement required 35 s of sustained evidence for both a weak selected
player and a superior challenger.

At each frame, the active set was the set of up to 11 players selected by this
procedure. Participation changes were inferred from the evolving set rather
than from an official substitution log, and a centroid estimate required at
least seven selected players with valid coordinates. Supplementary Methods S1,
Algorithm S1 and Supplementary Table S4 report the complete equations, weights
and update rules. A sensitivity analysis removed velocity coherence while
retaining the remaining reconstruction rules.

\subsection*{Collective coordinates, order and directional runs}

For active-player set $\mathcal A(t)$ with $N(t)=|\mathcal A(t)|$, the team
centroid was
\begin{equation}
    \bm X_{\mathrm c}(t)
    =
    \frac{1}{N(t)}
    \sum_{i\in\mathcal A(t)}
    \bm x_i(t).
    \label{eq:centroid_methods}
\end{equation}
Player position relative to the translating team was
$\bm r_i(t)=\bm x_i(t)-\bm X_{\mathrm c}(t)$, and collective order was measured
by the polarisation in \EqRef{eq:polarisation}. Figure~3 used global terciles
of the pooled run-level onset-order distribution. The cross-fitted Figure~4
and Figure~5 analyses instead used team-specific interval-state cutpoints
estimated only from the training match dates and applied unchanged to the
excluded group.

Order was summarised at the temporal scale required by each analysis. Figure~3
used mean polarisation over the first up to 3 s of a run, retaining shorter
runs, and Figure~3C used whole-run mean polarisation descriptively. Interval
models used current polarisation or its mean over the strictly preceding 5 s.
Figure~4B classified that preceding history by its modal state, excluding the
current interval, whereas Figure~5 used the current one-second state to
estimate transitions. Supplementary Methods S2 gives the exact definitions.

Directional runs were periods between successive heading changes greater than
$\theta_c=30^{\circ}$, using the same rule for centroid and player
trajectories. Paths were split at match-phase boundaries and timestamp gaps
greater than 2 s, and only paths containing at least 30 one-second observations
were segmented. Run duration was elapsed time and run length was cumulative
path length. An ending followed by another run in the same path was treated as
an observed termination; otherwise it was right-censored and contributed
exposure without an event. Supplementary Methods S2 reports the exact gap and
censoring rules.

\subsection*{Transport and duration statistics}

We characterised transport using completed-run distributions and mean-squared
displacement (MSD). Complementary cumulative distributions described run
duration and length, with mean-matched exponentials as memoryless,
constant-rate references. Completed centroid-run durations were compared with
a geometric memoryless baseline, log-normal, Weibull and power-law-based
alternatives, including the generalised-cutoff model
\begin{equation}
    p(\mathcal T=t)
    \propto
    t^{-\gamma}\exp\!\left[-(t/T_c)^{\beta}\right],
    \qquad t\geq1~\mathrm{s},
    \label{eq:generalised_cutoff_distribution}
\end{equation}
where $\gamma$ controls the broad early decay, $T_c$ the cutoff timescale and
$\beta$ the sharpness of the late attenuation; $\beta=1$ gives conventional
exponential truncation. Parameters were estimated by
maximum likelihood. Relative support was assessed from paired held-out
log-likelihood differences per run; absolute agreement was assessed using
parametric-bootstrap Kolmogorov--Smirnov statistics in which each simulated
dataset was refitted. Supplementary Methods S3 gives the full distributional
details.

MSD was calculated from all valid within-path pairs at each lag for centroid,
pitch-frame player and centroid-relative coordinates. Effective exponents were
estimated from log--log slopes over 5--30 s, after the initial 1--4 s
persistence regime. The decomposition in \EqRef{eq:msd_decomposition} used the
same displacement pairs and provided an exact signed decomposition of
pitch-frame player MSD. Supplementary Methods S3 reports the calculation and
bootstrap procedures.

\subsection*{Termination, polarisation and predictive validation}

Each centroid run was represented as a sequence of one-second intervals at
risk of termination. At each age, empirical termination probability was the
number of observed terminations divided by the number of runs still at risk.
Primary inference was restricted to 0--35 s because older risk sets were
sparse. The age-only, age-plus-order and full models were defined by
\EqRef{eq:inverse_age_hazard} and \EqRef{eq:full_hazard}. Continuous hazards
were integrated over each interval and fitted by Bernoulli maximum likelihood.

Association and prediction were evaluated separately. Logistic models tested
whether polarisation remained associated with termination after adjustment for
age and kinematics. The contemporaneous analysis used current polarisation and
mean active-player speed; the stricter lagged analysis used only the preceding
5-s history and adjusted for preceding active-player speed, centroid speed or
both. Nested held-out models then tested whether preceding polarisation added
predictive information beyond age and centroid kinematics, with recent
directional stability included as a further diagnostic adjustment. Predictors
were standardised within team using training-fold quantities. Supplementary
Methods S4 reports the model formulas and contrasts.

Generalisation was assessed using grouped leave-one-match-date-out
cross-fitting over 47 date groups. All records played on one date were withheld,
and thresholds, scaling quantities and parameters were estimated from the
remaining dates. Figure~4C supplied the observed held-out preceding-order
sequence as a time-varying covariate, testing transfer of the termination
relationship rather than forecasting future order. Figures~5B--C instead
propagated state composition from training-estimated birth distributions,
transition matrices and termination probabilities without replaying future
held-out states. This is internal out-of-sample validation within the same two
clubs, seasons and tracking system, not external validation. Predictions were
evaluated by log loss, Brier score and logistic calibration, defined in
Supplementary Methods S5.

\subsection*{State switching and survivor composition}

We modelled survivor composition through state switching and state-specific
termination. Current polarisation was discretised into fold-specific,
team-relative terciles. One-second transition matrices conditional on survival
were estimated in four age bands: 0--5, 5--10, 10--20 and 20--35 s. Starting
from the training-estimated state distribution at run birth, the age-banded
model propagated the live population using these transitions and
state-specific termination probabilities (\EqRef{eq:discrete_state_model}). A
stationary comparator used one transition matrix at all ages.

To separate the two processes, we removed one fitted component at a time. One
model retained age-banded switching but imposed common termination across
states; the other retained state-dependent termination but prevented
switching. Because each change alters the population on which subsequent
updates act, the resulting finite-horizon trajectories are not additive.
Supplementary Methods S6 describes the estimation and continuous-age
switching--selection decomposition.

\subsection*{Robustness and uncertainty}

We tested sensitivity to trajectory reconstruction and run segmentation using
a centred three-sample rolling median, low-speed turn guards of $0.25$ and
$0.50~\mathrm{m\,s^{-1}}$, and alternative turn thresholds. The
order--termination analyses were repeated with adjustment for speed and
directional stability, and active players were reconstructed without velocity
coherence. Uncertainty was estimated by bootstrap resampling at the match-date
or match--team--phase--source level, as appropriate, preserving dependence
within recording groups. Figure captions and Supplementary Methods S7 specify
the resampling unit and number of replicates. All reported intervals are 95\%
bootstrap intervals.

\subsection*{Ethics and data governance}

This study used anonymised, publicly available SoccerMon data and involved no
new participant recruitment, intervention or data collection. In the original
study, players provided written informed consent, including consent for open
publication, and identifiers were replaced before release
\cite{Midoglu2024}. The original collection was approved by the Norwegian
Privacy Data Protection Authority (reference 296155) and exempted from
additional approval by the relevant Regional Committees for Medical and Health
Research Ethics. We used only the released positional records and made no
attempt to identify players.


\section*{Data availability}

The SoccerMon dataset is publicly available through Zenodo under a Creative
Commons Attribution 4.0 International licence, with the permanent identifier
\url{https://doi.org/10.5281/zenodo.10033832}
\cite{SoccerMonZenodo}. The present study used the objective position records
from the 2020 and 2021 seasons. Derived, non-identifying source-data tables
underlying the figures and reported numerical results are available at
\url{https://github.com/framewave-lab/Athlelorien}.

\section*{Code availability}

Analysis code, frozen configuration files, and scripts used to generate the
figures and tables are available at
\url{https://github.com/framewave-lab/Athlelorien}.

\section*{Acknowledgements}

We thank the SoccerMon project team, the participating clubs and the players
for collecting and openly releasing the SoccerMon dataset used in this study. The authors
received no specific funding for this work.

\section*{Author contributions}

G.H.S., D.G., D.R. and S.K. contributed equally to study
conceptualisation, methodology, software development, data curation, formal
analysis, interpretation, visualisation, and drafting and revision of the
manuscript. All authors read and approved the final manuscript.

\section*{Competing interests}

The authors declare no competing interests.

\nolinenumbers


\clearpage
\setcounter{figure}{0}
\renewcommand{\thefigure}{S\arabic{figure}}
\renewcommand{\theHfigure}{S.\arabic{figure}}
\setcounter{table}{0}
\renewcommand{\thetable}{S\arabic{table}}
\renewcommand{\theHtable}{S.\arabic{table}}
\setcounter{algorithm}{0}
\renewcommand{\thealgorithm}{S\arabic{algorithm}}
\setcounter{equation}{0}
\renewcommand{\theequation}{S\arabic{equation}}
\renewcommand{\theHequation}{S.\arabic{equation}}
\renewcommand{\theHalgorithm}{S.\arabic{algorithm}}

\section*{Supplementary Information}

\begingroup
\singlespacing
\setlength{\parindent}{0pt}
\setlength{\parskip}{0.35em plus 0.08em minus 0.05em}
\captionsetup[figure]{font=footnotesize,skip=4pt}
\captionsetup[table]{font=footnotesize,skip=4pt}

The main Methods describe the operational definitions and inferential design.
This Supplement gives the implementation details, model formulas, parameter
values and robustness procedures needed to reproduce the analysis.

\subsection*{Supplementary Methods}
\subsubsection*{S1. Active-player reconstruction}

Pitch-calibrated coordinates were represented by signed depth from the
calibrated pitch-aligned rectangle,
\begin{equation}
 d_i(t)=\min\{x_i-x_{\min},\,x_{\max}-x_i,\,y_i-y_{\min},\,y_{\max}-y_i\}.
\end{equation}
Positive values lie inside the pitch. A preliminary geometry state became
active after 70 s continuously at least 6 m inside the boundary and returned
to bench after 110 s continuously off pitch. This state was one score
component rather than the final active-set label.

Recent path length and mean velocity were calculated over 45 one-second
samples. The initial XI was the 11 highest-scoring players over the first
inclusive 120 s, with seed score
\begin{align}
S_i={}&0.15\,\max(\mathrm{recent\ path}_i)
+0.02\sum I(d_i\geq6)
+0.01\sum I(d_i\geq0.1)\nonumber\\
&+0.03\,\mathrm{median}(d_i^+)
+0.02\sum I(\mathrm{geometry\ active}_i),
\end{align}
where $d_i^+=\max(d_i,0)$. At each later frame, candidates required finite
coordinates, $d_i\geq-30$ m, and either $d_i\geq-5$ m or membership of the
previous selected set.

For each candidate, velocity coherence $C_i\in[0,1]$ was the rescaled mean of
the seven largest finite pairwise cosine similarities between recent velocity
directions; headings below $0.25~\mathrm{m\,s^{-1}}$ were treated as undefined.
The base and hierarchical scores were
\begin{align}
L_i={}&0.10\,\mathrm{recent\ path}_i+0.025\,d_i^+
+0.25I(\mathrm{geometry\ active}_i)+0.20I(d_i\geq6)\nonumber\\
&+0.35I(i\in\mathrm{seed\ XI})+1.50C_i
+0.35\,\operatorname{clip}\!\left(\frac{d_i+5}{17},0,1\right),\\
H_i={}&L_i+2.50I(i\in A(t^-)).
\end{align}
Vacancies were filled by the largest $H_i$. When 11 players were already
selected, a player accumulated weak evidence while $L_i<1.6$, and a challenger
accumulated replacement evidence while exceeding the weakest selected score by
at least 0.75. Replacement required 35 s of both conditions. The active set was
capped at 11 and centroid estimates required at least seven selected players.
Exact score ties inherited the existing input order. Algorithm~S1 summarises
the production update and Supplementary Table~S4 lists the frozen parameters.
The no-coherence sensitivity set the coefficient of $C_i$ to zero while
retaining all other rules.

\medskip
\refstepcounter{algorithm}
\noindent\textbf{Algorithm \thealgorithm. Active-player reconstruction used
for the production cache.}
\label{alg:supp_active_player}
\begin{algorithmic}[1]
\Require Pitch-calibrated player rows and timestamps.
\Ensure One label per row: \texttt{active}, \texttt{bench} or
\texttt{rejected\_hoverer}.
\State Compute signed depth $d_i$, preliminary 70/110-s geometry states and
45-s recent-motion features.
\State Seed $A_{\mathrm{prev}}$ with the 11 largest $S_i$ over the first 120 s.
\For{each frame $t$ in time order}
    \State Form candidates with finite coordinates, $d_i\geq-30$ m, and
    $d_i\geq-5$ m or $i\in A_{\mathrm{prev}}$.
    \State Compute $C_i$, $L_i$ and $H_i$ from Eqs.~(S3)--(S4).
    \State Retain previous selections still present among candidates.
    \While{$|A(t)|<11$ and an unselected candidate remains}
        \State Add the candidate with largest $H_i$.
    \EndWhile
    \State Accumulate weak time for selected players with $L_i<1.6$ and
    challenger time for non-selected players with
    $L_i\geq\min_{j\in A(t)}L_j+0.75$.
    \While{a weak selected player and challenger both have at least 35 s evidence}
        \State Replace the weakest selected player with the strongest ready challenger.
    \EndWhile
    \State Keep at most 11 players, assign row labels, and set
    $A_{\mathrm{prev}}=A(t)$.
\EndFor
\State Retain centroid frames with at least seven active players.
\end{algorithmic}
\medskip

\subsubsection*{S2. Temporal processing, order summaries and run construction}

Timestamps were reconstructed from recording date and logged time of day using
the Norwegian daylight-saving offset. Within each player track, velocity was
calculated by backward difference,
\begin{equation}
\bm v_i(t)=\frac{\bm x_i(t)-\bm x_i(t-\Delta t)}{\Delta t},
\qquad \Delta t\simeq1~\mathrm{s},
\end{equation}
using the observed timestamp difference. Velocity was undefined at the first
sample of a continuous path. Zero-speed and otherwise undefined directions
were excluded from polarisation, and mean active-player speed was the mean of
valid player speeds at a frame.

Figure~3 cutpoints were global terciles of onset order across the pooled
centroid runs. Figure~3 onset order was the mean of all available one-second
polarisation samples from run birth through the earlier of run end or 3 s. Figure~3C used whole-run mean polarisation only for
descriptive visualisation. The contemporaneous logistic model used current
interval polarisation. Figure~4B used the modal categorical state among
available intervals in the strictly preceding 5 s, excluding the current
interval; ties were resolved with the most recent prior state. The lagged
adjustment analysis used mean continuous polarisation over the same history.
Figure~5 used current one-second polarisation discretised with fold-specific
training thresholds.

Continuous source trajectories were split at match-phase boundaries and at
timestamp gaps strictly greater than 2 s. Paths were retained only when they
contained at least 30 one-second samples before segmentation. Consecutive
displacement headings were compared by circular difference, and a new run
began when the change exceeded $30^{\circ}$. Displacement vectors with norm at
most $10^{-6}$ m were skipped in turn evaluation and did not initiate a new
run. Retained runs contained at least two samples and positive arc length;
interval analyses required duration of at least 1 s.

Run duration was elapsed time between first and final samples, and path length
was
\begin{equation}
L_k=\sum_j\left|\bm X_{\mathrm c}(t_{j+1})-\bm X_{\mathrm c}(t_j)\right|.
\end{equation}
An ending was classified as an observed directional termination when the next
run in the same track began between $-0.25$ and 2.0 s relative to the current
ending. Otherwise it was right-censored. Censored runs contributed exposure
through their final observed interval but not an event.

\subsubsection*{S3. Duration-family fitting and transport statistics}

For non-negative run measure $W$ with empirical mean $\bar W$, the descriptive
mean-matched exponential reference was
\begin{equation}
S_{\exp}(w)=\exp\!\left(-\frac{w}{\bar W}\right).
\end{equation}
Centroid and player references were matched separately. These curves were
descriptive and were not used to select a unique tail family.

Duration families were fitted on integer support $t\geq1$ s; $t\geq4$ s was
examined as a tail sensitivity. The generalised-cutoff probability mass was
\begin{equation}
p(\mathcal T=t)=
\frac{t^{-\gamma}\exp[-(t/T_c)^\beta]}
{\sum_{u=1}^{10{,}000}u^{-\gamma}\exp[-(u/T_c)^\beta]}.
\end{equation}
Pure and conventionally truncated power laws were normalised on the same grid.
For log-normal and Weibull models, the mass assigned to integer second $t$ was
the corresponding continuous probability integrated over
$[\max(t-0.5,0.5),t+0.5)$. The geometric model---the discrete-time analogue
of a memoryless exponential---used $p(t)=\rho(1-\rho)^{t-1}$. Parameters were estimated by maximum likelihood on
training date groups and evaluated on the excluded date group. Model ranking
used paired held-out log-likelihood contrasts per run. Absolute fit used
parametric-bootstrap Kolmogorov--Smirnov statistics, refitting each simulated
dataset before calculating its statistic.

For coordinate $\bm y(t)$, MSD was
\begin{equation}
M_y(\tau)=\left\langle|\bm y(t+\tau)-\bm y(t)|^2\right\rangle,
\end{equation}
using all valid within-path pairs. The principal exponent was estimated from
\begin{equation}
\log M_y(\tau)=c+\alpha_{\mathrm{MSD}}\log\tau
\end{equation}
over 5--30 s. The 1--4 s regime was descriptive and excluded from the primary
fit. Centroid, pitch-frame player and centroid-relative player MSDs used the
same lag grid and displacement pairs.

\subsubsection*{S4. Termination and polarisation adjustment}

Each centroid run was represented by one-second intervals with run age,
exposure, event status, order and speed. In age bin $j$, empirical termination
probability was $d_j/n_j$, and the corresponding survivor was
\begin{equation}
\widehat S(a)=\prod_{a_j\leq a}\left(1-\frac{d_j}{n_j}\right).
\end{equation}
For interval $[a,a+\Delta a)$, predicted termination probability was
\begin{equation}
h_k(a)=1-\exp\!\left[-\int_a^{a+\Delta a}\lambda_k(u)\,\dd u\right].
\end{equation}
Parameters were estimated by maximising
\begin{equation}
\ell=\sum_{k,j}\left[y_{kj}\log h_{kj}+(1-y_{kj})\log(1-h_{kj})\right].
\end{equation}

Across folds, the age-only model had median
$(\mu,a_0)=(1.33,2.55~\mathrm{s})$, while the age-plus-order model had
$(\mu,a_0)=(1.92,4.77~\mathrm{s})$. The later-age contribution had median
parameters $q=0.104~\mathrm{s}^{-1}$, $t_c=20~\mathrm{s}$ and
$\tau_q=5~\mathrm{s}$.

The contemporaneous speed-adjusted model was
\begin{equation}
\operatorname{logit}\Prob(Y_i=1)=
\alpha_0+\beta_p z_{p,i}+\beta_a\log(1+a_i)+\beta_v z_{v,i},
\end{equation}
where polarisation and mean active-player speed were standardised within team.
Right-censored terminal intervals were not counted as observed events.
Speed-stratum cutpoints and standardisation quantities were estimated from the
training groups within each fold.

For the lagged analysis,
\begin{equation}
p_{\mathrm{prev5}}(a)=\frac{1}{|\mathcal H_a|}\sum_{u\in\mathcal H_a}p(u),
\qquad
\mathcal H_a=\{u:\max(0,a-5)\leq u<a\}.
\end{equation}
Only prior valid intervals from the same run were used; missing history was not
interpolated. The first interval therefore lacked prior history and was
excluded. Lagged analyses contained 285,528 intervals, 40,505 terminations and
114 right-censored terminal intervals retained as non-events.

Nested models contained age alone (M0); age and preceding centroid speed (M1);
age and preceding polarisation (M2); age, preceding centroid speed and
preceding polarisation (M3); age, preceding centroid speed and preceding
centroid directional stability (M4); and M4 plus preceding polarisation (M5).
Directional stability was the resultant length of preceding centroid headings.
The primary incremental contrast was M3 minus M1; M5 minus M4 additionally
conditioned on recent directional persistence. Coefficient intervals used 500
match-date-group bootstrap refits, and predictive contrasts used 2,000 paired
match-date-group bootstrap resamples of completed out-of-fold predictions.

\subsubsection*{S5. Cross-validation and predictive scores}

Grouped leave-one-match-date-out cross-fitting used 47 folds. Each fold removed
all records associated with one date, including both tracked teams in the four
head-to-head fixtures and both distinct fixtures on the 15 dates when the clubs
played separate opponents. Team-specific thresholds, predictor scaling,
hazard parameters, birth distributions and transition matrices were estimated
from the remaining dates and applied unchanged to the held-out group.

Figure~4C used the observed held-out preceding-order state as a time-varying
covariate. Figures~5B--C instead propagated live-state masses from
training-estimated quantities without future held-out states. Fold predictions
were pooled only after every interval or state trajectory had received an
out-of-fold prediction.

For outcome $y_i\in\{0,1\}$ and out-of-fold probability $\widehat p_i$,
\begin{align}
\mathrm{LogLoss}&=-\frac{1}{N}\sum_i
[y_i\log\widehat p_i+(1-y_i)\log(1-\widehat p_i)],\\
\mathrm{Brier}&=\frac{1}{N}\sum_i(\widehat p_i-y_i)^2.
\end{align}
Calibration intercept and slope were obtained by logistic recalibration of
pooled outcomes against the logit of predicted probability; ideal values are
zero and one. Total held-out log likelihood was retained in source data but
omitted from the main predictive table because it is redundant with mean log
loss for a fixed evaluation set.

\subsubsection*{S6. State-transition and process-removal models}

For age band $b(j)$, the transition probability conditional on survival was
\begin{equation}
P_{sr,j}=\Prob(Z_{j+1}=r\mid Z_j=s,\ \text{survival of interval }j),
\qquad \bm P_j=\bm P^{(b(j))}.
\end{equation}
Death and censoring were not next states. With training-estimated birth
distribution $\bm u_0=\bm\pi_0$ and state-specific termination probability
$d_{s,j}$,
\begin{equation}
\bm u_{j+1}=\bm A_j^{\mathsf T}\bm u_j,
\qquad A_{sr,j}=(1-d_{s,j})P_{sr,j}.
\end{equation}
The stationary comparator replaced the four age-banded matrices by one matrix.

The formal continuous-age interpretation used birth boundary
\begin{equation}
\bm n(0,t)=B(t)\bm\pi_0,
\end{equation}
where $B(t)$ is the rate of run formation. For the high-order component,
\begin{align}
\frac{\dd c_H}{\dd a}={}&
\kappa_{LH}c_L+\kappa_{MH}c_M
-(\kappa_{HL}+\kappa_{HM})c_H
+[\overline\lambda-\lambda_H]c_H,
\end{align}
with age dependence suppressed for readability. This representation was used
for interpretation, not fitted as a separate smooth generator.

For the exact one-step decomposition, let $r_{s,j}=1-d_{s,j}$. Differential
termination first gives
\begin{equation}
\widetilde c_{s,j}=\frac{r_{s,j}c_{s,j}}{\sum_r r_{r,j}c_{r,j}},
\end{equation}
and switching then gives
$\bm c_{j+1}=\bm P_j^{\mathsf T}\widetilde{\bm c}_j$. Thus
\begin{equation}
\bm c_{j+1}-\bm c_j=
(\widetilde{\bm c}_j-\bm c_j)
+(\bm c_{j+1}-\widetilde{\bm c}_j).
\end{equation}
The first term is selection by differential termination and the second is
redistribution by switching.

The switching-with-common-termination model retained the transition matrices
but replaced state-specific inverse-age multipliers by
$m_{\mathrm{common}}=\bm\pi_0^{\mathsf T}\bm m$. The common late contribution
was retained. The termination-without-switching model retained state-specific
termination probabilities and replaced every transition matrix by the
identity. Both variants were propagated recursively within each fold and did
not use future held-out states.

\subsubsection*{S7. Robustness and uncertainty}

The primary run segmentation was compared with a centred three-sample rolling
median applied separately to centroid $x$ and $y$, and with low-speed turn
guards at $0.25$ and $0.50~\mathrm{m\,s^{-1}}$. Under a guard, a candidate turn
was suppressed when either adjacent centroid step was slower than the
threshold; the candidate turn was merged into the continuing run and exposure
was retained. Run counts, censoring, duration quantiles, survivors,
termination probabilities and inverse-age parameters were recomputed for each
definition. Supplementary Figure~S2 used 500 match-date-group bootstrap
resamples.

Dependence was preserved by resampling match-date groups or
match--team--phase--source clusters, where source denotes the original
SoccerMon trajectory source from which a continuous phase-specific path was
constructed. Figure~2 used 200 cluster-bootstrap replicates. Figure~3,
coefficient estimates and segmentation analyses used 500 grouped replicates.
Incremental predictive contrasts used 2,000 paired resamples of completed
out-of-fold predictions. All intervals are 95\% bootstrap intervals.

\clearpage
\section*{Supplementary Figures}

\noindent\begin{minipage}{\textwidth}
\centering
\includegraphics[width=0.96\textwidth,keepaspectratio]{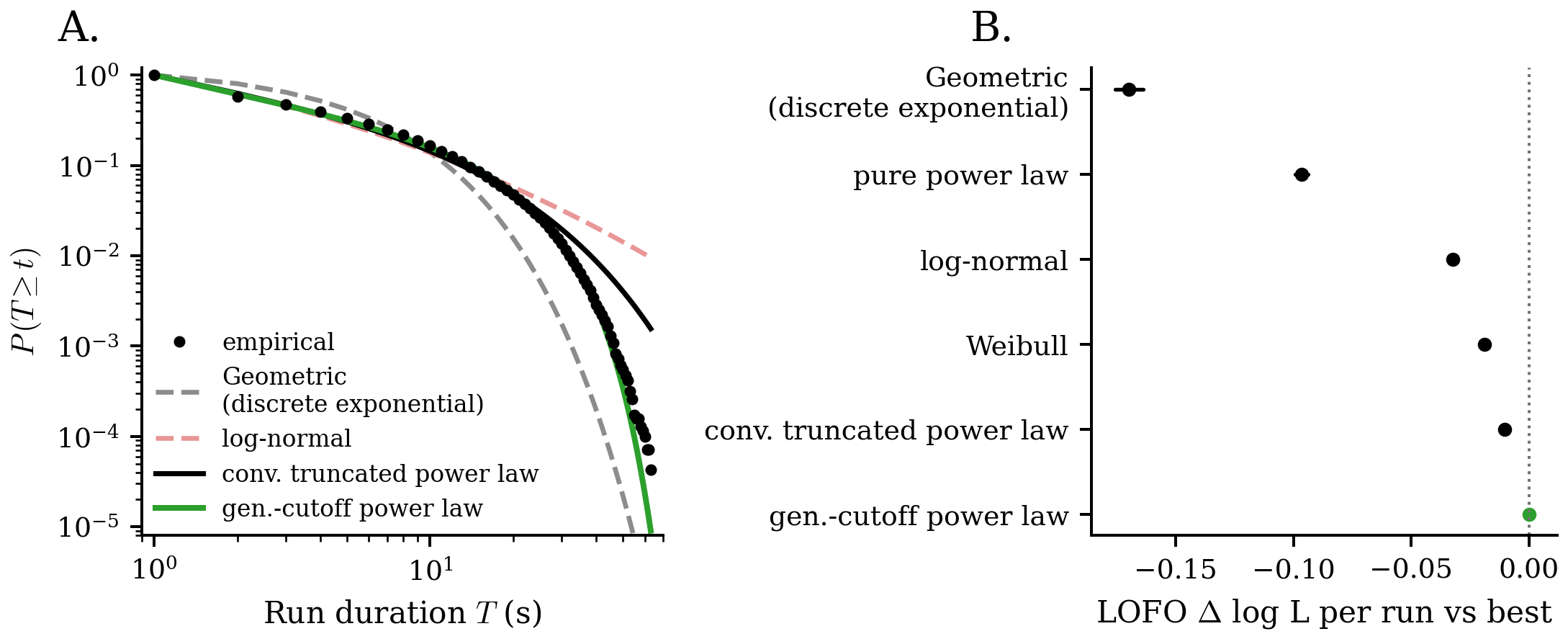}
\captionof{figure}{\textbf{Duration-family comparison.}
\textbf{A}, Empirical completed centroid-run duration survivor and selected
candidate fits over the full observed support. The green curve is the
generalised-cutoff power law; the black, red and grey curves show the
conventional truncated power law, log-normal and geometric models, respectively.
The geometric model is the discrete-time analogue of a memoryless exponential.
\textbf{B}, Grouped leave-one-match-date-out (LOFO) log-likelihood
difference per run, defined as comparator minus generalised-cutoff log
likelihood. Negative values therefore favour the generalised-cutoff model;
horizontal intervals show the 95\% uncertainty in the held-out contrast. The
comparison ranks candidate families relatively; parametric-bootstrap
goodness-of-fit tests detected deviations for every family tested.}
\label{fig:supp_duration}
\end{minipage}

\vspace{1.0em}

\noindent\begin{minipage}{\textwidth}
\centering
\includegraphics[width=0.96\textwidth,keepaspectratio]{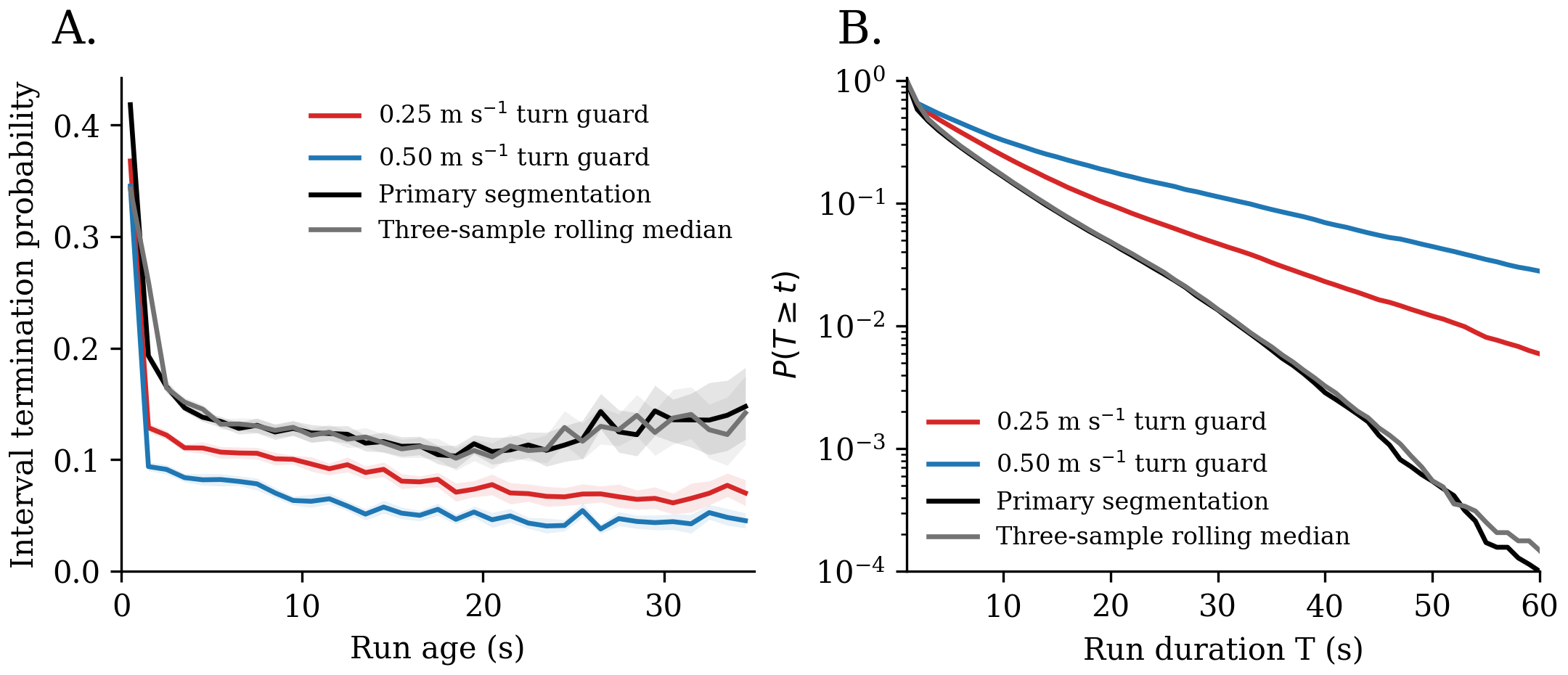}
\captionof{figure}{\textbf{Robustness of age-dependent termination to run segmentation.}
\textbf{A}, Empirical one-second termination probability under the primary
$30^{\circ}$ segmentation, a centred three-sample rolling median and low-speed
turn guards of $0.25$ and $0.50~\mathrm{m\,s^{-1}}$. Under the guarded
definitions, low-speed candidate turns were merged into the continuing run
rather than omitted. Ribbons show 95\% intervals from 500 match-date-group
bootstrap resamples. \textbf{B}, Corresponding duration survivor curves. The
rolling-median result closely follows the primary analysis. Low-speed guards
produce fewer, longer runs and lower absolute termination probabilities, but
preserve the high onset risk, subsequent decline and later flattening or
increase.}
\label{fig:supp_segmentation}
\end{minipage}

\clearpage
\section*{Supplementary Tables}

\footnotesize
\renewcommand{\arraystretch}{0.92}
\setlength{\tabcolsep}{3pt}
\begin{longtable}{@{}p{0.23\linewidth}p{0.14\linewidth}p{0.55\linewidth}@{}}
\caption{\textbf{Dataset and processing summary.}}
\label{tab:supp_dataset}\\
\toprule
Quantity & Value & Definition\\
\midrule
\endfirsthead
\caption[]{\textbf{Table S1 continued.}}\\
\toprule
Quantity & Value & Definition\\
\midrule
\endhead
\multicolumn{3}{l}{\textbf{Dataset}}\\
Competitive fixtures & 62 & Distinct Norwegian Toppserien matches.\\
Match dates & 47 & Date groups used for grouped cross-validation and bootstrap.\\
Tracked team-match records & 66 & One SoccerMon-tracked club in one competitive fixture.\\
Seasons & 2 & 2020 and 2021 Norwegian Toppserien seasons.\\
Data source & SoccerMon & Public GPS position records for Rosenborg and V\aa lerenga.\\
Clubs & 2 & Rosenborg and V\aa lerenga.\\
Centroid runs & 69,802 & Directional centroid runs.\\
Observed terminations & 69,670 & Completed directional run endings.\\
Right-censored endings & 132 & Run endings without an observed subsequent directional change.\\
One-second risk intervals & 355,330 & Intervals included in termination models.\\
Player runs & 1,192,715 & Player-level directional runs.\\
\addlinespace
\multicolumn{3}{l}{\textbf{Temporal processing}}\\
Native / analysis frequency & 10 Hz / 1 Hz & First timestamp-sorted observation retained in each one-second bin.\\
Empty bins & Not filled & No interpolation or forward filling.\\
Temporal smoothing & None & No smoothing before velocity calculation.\\
Velocity & Backward difference & Per-player displacement divided by elapsed time.\\
Zero-speed headings & Excluded & Undefined unit headings omitted from polarisation.\\
Turn-vector guard & $\leq10^{-6}$ m skipped & Near-zero vectors did not create turns.\\
\addlinespace
\multicolumn{3}{l}{\textbf{Path, active-player and run settings}}\\
Gap threshold & $>2$ s & Larger timestamp gaps split continuous paths.\\
Minimum path & 30 samples & Shorter player or centroid paths were excluded.\\
Active-set cap / centroid minimum & 11 / 7 & Maximum selected players and minimum required for a centroid frame.\\
Active-zone interior depth & 6 m & Preliminary geometry feature requiring position at least 6 m inside the pitch boundary.\\
Soft / hard candidate margins & 5 m / 30 m & Soft outside-pitch eligibility margin and hard coordinate-sanity margin.\\
Geometry-state persistence & 70 s / 110 s & Preliminary active-zone activation and off-pitch demotion periods.\\
Final replacement confirmation & 35 s / 35 s & Weak selected-player and superior challenger evidence required for replacement.\\
Initial-XI seed window & First 120 s & Inclusive window used to initialise the selected set.\\
Primary turn threshold & $30^{\circ}$ & Directional-run segmentation.\\
Minimum analysed run & 1 s & Entry threshold for interval analysis.\\
Onset-order window & First up to 3 s & Figure~3; shorter runs use all available samples.\\
Figure~4 recent state & Preceding 5 s & Modal categorical state; current interval excluded.\\
Figure~5 current state & Current 1-s interval & Fold-thresholded current polarisation.\\
State age bands & 0--5, 5--10, 10--20, 20--35 s & Age-banded transition matrices.\\
\bottomrule
\end{longtable}

\clearpage
\begin{longtable}{@{}p{0.36\linewidth}p{0.11\linewidth}p{0.17\linewidth}p{0.28\linewidth}@{}}
\caption{\textbf{Model comparison and robustness estimates.} Duration-model
contrasts are comparator minus generalised-cutoff held-out log likelihood per
run. Odds ratios are per one-standard-deviation increase in within-team
polarisation. Predictive-score differences are larger model minus smaller
model, so negative log-loss and Brier-score values favour adding
polarisation.}
\label{tab:supp_model_robustness}\\
\toprule
Analysis & Estimate & 95\% interval & Sample or diagnostic\\
\midrule
\endfirsthead
\caption[]{\textbf{Table S2 continued.}}\\
\toprule
Analysis & Estimate & 95\% interval & Sample or diagnostic\\
\midrule
\endhead
\multicolumn{4}{l}{\textbf{A. Duration-family comparison}}\\
Geometric (discrete exponential) & $-0.1698$ & $[-0.1757,-0.1636]$ & rank 6; KS $0.222$\\
Pure power law & $-0.0964$ & $[-0.0991,-0.0937]$ & rank 5; KS $0.091$\\
Log-normal & $-0.0325$ & $[-0.0339,-0.0311]$ & rank 4; KS $0.047$\\
Weibull & $-0.0189$ & $[-0.0201,-0.0178]$ & rank 3; KS $0.053$\\
Conventional truncated power law & $-0.0103$ & $[-0.0111,-0.0095]$ & rank 2; KS $0.047$\\
Generalised-cutoff power law & --- & --- & rank 1; KS $0.033$\\
Generalised-cutoff parameters & $\gamma=1.326$ & --- & $T_c=32.3$ s; $\beta=2.782$\\
\addlinespace
\multicolumn{4}{l}{\textbf{B. Current-interval polarisation (contemporaneous)}}\\
Adjusted for age and mean active-player speed & $0.802$ & $[0.792,0.814]$ & 355,330 / 69,670 intervals/events\\
Slow speed stratum & $0.817$ & $[0.802,0.832]$ & 118,442 / 28,508\\
Middle speed stratum & $0.800$ & $[0.781,0.819]$ & 118,420 / 23,335\\
Fast speed stratum & $0.712$ & $[0.690,0.737]$ & 118,468 / 17,827\\
Reconstruction excluding velocity coherence & $0.814$ & $[0.802,0.826]$ & 350,231 / 68,437\\
\addlinespace
\multicolumn{4}{l}{\textbf{C. Preceding-five-second polarisation}}\\
Adjusted for preceding mean active-player speed & $0.816$ & $[0.803,0.834]$ & log loss $0.404$; Brier $0.121$\\
Adjusted for preceding centroid speed & $0.945$ & $[0.926,0.964]$ & log loss $0.403$; Brier $0.120$\\
Both speed measures (collinearity diagnostic) & $0.979$ & $[0.960,0.999]$ & $r=0.918$; maximum VIF $8.0$\\
\addlinespace
\multicolumn{4}{l}{\textbf{D. Turn-threshold sensitivity}}\\
$20^{\circ}$ threshold & $0.836$ & $[0.826,0.845]$ & 106,223 runs; median 1 s; 95th 12 s\\
$30^{\circ}$ threshold & $0.805$ & $[0.793,0.817]$ & 69,802 runs; median 2 s; 95th 19 s\\
$40^{\circ}$ threshold & $0.774$ & $[0.760,0.786]$ & 49,579 runs; median 4 s; 95th 26 s\\
\addlinespace
\multicolumn{4}{l}{\textbf{E. Incremental held-out prediction from preceding polarisation}}\\
M3 vs M1: $\Delta$ log loss ($10^{-4}$) & $-0.768$ & $[-1.337,-0.215]$ & 285,528 intervals; 25/47 date groups favour M3\\
M3 vs M1: $\Delta$ Brier score ($10^{-5}$) & $-1.278$ & $[-2.764,0.146]$ & 21/47 date groups favour M3\\
M5 vs M4: $\Delta$ log loss ($10^{-4}$) & $-0.665$ & $[-1.187,-0.158]$ & 285,528 intervals; 24/47 date groups favour M5\\
M5 vs M4: $\Delta$ Brier score ($10^{-5}$) & $-0.936$ & $[-2.290,0.372]$ & 22/47 date groups favour M5\\
M5 lagged-polarisation odds ratio & $0.948$ & $[0.930,0.968]$ & age + speed + directional stability; 114 censored intervals retained\\
\addlinespace
\multicolumn{4}{l}{\textbf{F. Segmentation robustness}}\\
Primary segmentation & 69,802 runs & --- & median 2 s; 95th 19 s; hazard minimum 18.5 s\\
$0.25~\mathrm{m\,s^{-1}}$ turn guard & 47,522 runs & --- & median 3 s; 95th 29 s; hazard minimum 30.5 s\\
$0.50~\mathrm{m\,s^{-1}}$ turn guard & 31,067 runs & --- & median 4 s; 95th 47 s; hazard minimum 26.5 s\\
Three-sample rolling median & 67,501 runs & --- & median 2 s; 95th 19 s; hazard minimum 18.5 s\\
\bottomrule
\end{longtable}
\normalsize

\noindent\footnotesize KS denotes the Kolmogorov--Smirnov statistic; smaller
values indicate a smaller maximum discrepancy between the fitted and empirical
distributions. In the incremental prediction analysis, M1 contains age and
preceding centroid speed; M3 adds preceding polarisation; M4 contains age,
preceding centroid speed and preceding directional stability; and M5 adds
preceding polarisation to M4. All candidate duration families showed detectable
absolute goodness-of-fit deviations. Polarisation coefficients quantify
associations and do not identify causal effects. Coefficient and segmentation
intervals used 500 match-date-group bootstrap replicates; incremental
prediction intervals used 2,000 paired match-date-group resamples of completed
out-of-fold predictions. The joint-speed model is reported only as a
collinearity diagnostic.

\clearpage

\noindent\begin{minipage}{\textwidth}
\centering
\captionof{table}{\textbf{Direct pooled age-banded transition matrices.}
Probabilities condition on survival through the one-second interval. Rows
denote the current state and columns the next state.}
\label{tab:supp_transitions}
\vspace{0.3em}
\begin{tabular}{llrrr}
\toprule
Age band & Current state & Next Low & Next Mid & Next High\\
\midrule
0--5 s & Low  & 0.635 & 0.306 & 0.059\\
       & Mid  & 0.140 & 0.548 & 0.312\\
       & High & 0.014 & 0.166 & 0.820\\
5--10 s & Low  & 0.747 & 0.234 & 0.018\\
        & Mid  & 0.155 & 0.633 & 0.212\\
        & High & 0.009 & 0.179 & 0.812\\
10--20 s & Low  & 0.782 & 0.206 & 0.012\\
         & Mid  & 0.181 & 0.642 & 0.177\\
         & High & 0.011 & 0.212 & 0.777\\
20--35 s & Low  & 0.801 & 0.190 & 0.010\\
         & Mid  & 0.237 & 0.616 & 0.147\\
         & High & 0.019 & 0.249 & 0.732\\
\bottomrule
\end{tabular}

\vspace{0.35em}
\raggedright\footnotesize
Termination and censoring are not next states. Unrounded rows sum to one.
Figure~5A uses these pooled descriptive matrices; Figures~5B--C use training
matrices estimated separately within each leave-one-match-date-out fold.
\end{minipage}

\vspace{1.2em}

\noindent\begin{minipage}{\textwidth}
\centering
\captionof{table}{\textbf{Frozen active-player reconstruction parameters.}
Values were recovered from the production implementation.}
\label{tab:supp_active_parameters}
\vspace{0.3em}
\begin{tabularx}{\textwidth}{@{}p{0.28\textwidth}p{0.18\textwidth}X@{}}
\toprule
Parameter or rule & Value & Role\\
\midrule
Production method & sticky hierarchical active XI & Final active-player selector\\
Active-zone depth & 6.0 m & Preliminary geometry feature requiring position inside every pitch boundary\\
Geometry activation / demotion & 70 s / 110 s & Persistence for preliminary geometry-state transitions\\
Soft / hard outside margins & 5.0 m / 30.0 m & Candidate eligibility and coordinate-sanity limits\\
Initial seed window / size & 120 s / 11 players & Initial selected set\\
Recent-motion window & 45 samples & Path and velocity features at 1 Hz\\
Weak-score threshold & 1.6 & Starts weak selected-player counter\\
Switch margin & 0.75 & Challenger advantage over weakest selected base score\\
Weak / challenger confirmation & 35 s / 35 s & Evidence required before replacement\\
Movement weight & 0.10 & Weight on recent path length\\
Pitch-depth weight & 0.025 & Weight on non-negative pitch depth\\
Geometry-active bonus & 0.25 & Bonus for preliminary active geometry state\\
Active-zone bonus & 0.20 & Bonus for $d_i\geq6$ m\\
Seed membership bonus & 0.35 & Bonus for initial-XI membership\\
Velocity-coherence weight & 1.50 & Weight on directional coherence\\
Velocity neighbours / speed floor & 7 / $0.25~\mathrm{m\,s^{-1}}$ & Coherence calculation\\
Position-score weight & 0.35 & Weight on clipped position score $(d_i+5)/17$\\
Continuity bonus & 2.50 & Hierarchical-score bonus for previous selection\\
Maximum active players & 11 & Framewise selected-set cap\\
Minimum players for centroid & 7 & Centroid frame-retention threshold\\
Exact score ties & Existing input order & No explicit secondary code-level tie-breaker\\
\bottomrule
\end{tabularx}
\end{minipage}

\renewcommand{\arraystretch}{1}
\normalsize
\endgroup
\end{document}